\documentstyle[10pt,epsf,epsfig,dp_delphititle,amssymb]{dp_delphi}
\makeindex
\def\DpPaperGroup{EP}
\def\DpPaperRef{2001-011}
\def\DpDate{Revised 18 January 2001}
\def\DpAuthors{DELPHI Collaboration}
\def\DpSubmit{(Accepted by Phys.Lett.B)}
\def\DpTitle{{Update of the
 search for supersymmetric particles in scenarios with \\
Gravitino LSP and Sleptons NLSP 
}}
\def\DpComment{ }
\def\DpEMail{ }
\begin{document}
%%%%%%%%%%%%%%%%%%%%%%%%%% They are a problem with Coll.Sty ?
\makeatletter
% Collapse citation numbers to ranges.  Non-numeric and undefined labels
% are handled.  No sorting is done.  E.g., 1,3,2,3,4,5,foo,1,2,3,?,4,5
% gives 1,3,2-5,foo,1-3,?,4,5
\newcount\@tempcntc
\def\@citex[#1]#2{\if@filesw\immediate\write\@auxout{\string\citation{#2}}\fi
  \@tempcnta\z@\@tempcntb\m@ne\def\@citea{}\@cite{\@for\@citeb:=#2\do
    {\@ifundefined
       {b@\@citeb}{\@citeo\@tempcntb\m@ne\@citea\def\@citea{,}{\bf ?}\@warning
       {Citation `\@citeb' on page \thepage \space undefined}}%
    {\setbox\z@\hbox{\global\@tempcntc0\csname b@\@citeb\endcsname\relax}%
     \ifnum\@tempcntc=\z@ \@citeo\@tempcntb\m@ne
       \@citea\def\@citea{,}\hbox{\csname b@\@citeb\endcsname}%
     \else
      \advance\@tempcntb\@ne
      \ifnum\@tempcntb=\@tempcntc
      \else\advance\@tempcntb\m@ne\@citeo
      \@tempcnta\@tempcntc\@tempcntb\@tempcntc\fi\fi}}\@citeo}{#1}}
\def\@citeo{\ifnum\@tempcnta>\@tempcntb\else\@citea\def\@citea{,}%
  \ifnum\@tempcnta=\@tempcntb\the\@tempcnta\else
   {\advance\@tempcnta\@ne\ifnum\@tempcnta=\@tempcntb \else \def\@citea{--}\fi
    \advance\@tempcnta\m@ne\the\@tempcnta\@citea\the\@tempcntb}\fi\fi}
 
\makeatother
%%%%%%%%%%%%%%%%%%%%%%%%%% ??????????????????????????????????
% Generate the title page
\begin{titlepage}
\pagenumbering{roman}
\CERNpreprint{\DpPaperGroup}{\DpPaperRef} % Reference of the paper
\date{{\small\DpDate}} % Date of the paper
\title{\DpTitle} % Title of the paper
\address{\DpAuthors} % General name of the author(s)
\begin{shortabs} % Start the abstract
\noindent
An update of the search for sleptons, neutralinos and charginos 
in the context of scenarios where the lightest 
supersymmetric particle is the gravitino and the next-to-lightest 
supersymmetric particle is a slepton, is presented, together with the 
update of the search for heavy stable charged particles in light 
gravitino scenarios and Minimal Supersymmetric Standard Models. 
Data collected in 1999 with the DELPHI detector at centre-of-mass energies
around 192, 196, 200 and 202 GeV were analysed.  No evidence 
for the production of these supersymmetric particles was found. 
Hence, new mass limits were derived at 95\% confidence level.

%%% Local Variables: 
%%% mode: latex
%%% TeX-master: t
%%% End: 
\end{shortabs}
\vfill
\begin{center}
\DpSubmit \ \\ % Horrible hack to allow to have DpSubmit empty
\DpComment \ \\
\DpEMail \ \\
\end{center}
\vfill
\clearpage
\vfill
\clearpage
\begingroup
% Commands to process the author names
%
\newcommand{\DpName}[2]{\hbox{#1$^{\ref{#2}}$},\hfill}
\newcommand{\DpNameTwo}[3]{\hbox{#1$^{\ref{#2},\ref{#3}}$},\hfill}
\newcommand{\DpNameThree}[4]{\hbox{#1$^{\ref{#2},\ref{#3},\ref{#4}}$},\hfill}
\newskip\Bigfill \Bigfill = 0pt plus 1000fill
\newcommand{\DpNameLast}[2]{\hbox{#1$^{\ref{#2}}$}\hspace{\Bigfill}}
\small
\noindent
\DpName{P.Abreu}{LIP}
\DpName{W.Adam}{VIENNA}
\DpName{T.Adye}{RAL}
\DpName{P.Adzic}{DEMOKRITOS}
\DpName{Z.Albrecht}{KARLSRUHE}
\DpName{T.Alderweireld}{AIM}
\DpName{G.D.Alekseev}{JINR}
\DpName{R.Alemany}{CERN}
\DpName{T.Allmendinger}{KARLSRUHE}
\DpName{P.P.Allport}{LIVERPOOL}
% ...
\normalsize
\endgroup
\titlefoot{Department of Physics and Astronomy, Iowa State
     University, Ames IA 50011-3160, USA
    \label{AMES}}
\titlefoot{Physics Department, Univ. Instelling Antwerpen,
     Universiteitsplein 1, B-2610 Antwerpen, Belgium \\
     \indent~~and IIHE, ULB-VUB,
     Pleinlaan 2, B-1050 Brussels, Belgium \\
     \indent~~and Facult\'e des Sciences,
     Univ. de l'Etat Mons, Av. Maistriau 19, B-7000 Mons, Belgium
    \label{AIM}}
\titlefoot{Physics Laboratory, University of Athens, Solonos Str.
     104, GR-10680 Athens, Greece
    \label{ATHENS}}
\titlefoot{Department of Physics, University of Bergen,
     All\'egaten 55, NO-5007 Bergen, Norway
    \label{BERGEN}}
\titlefoot{Dipartimento di Fisica, Universit\`a di Bologna and INFN,
     Via Irnerio 46, IT-40126 Bologna, Italy
    \label{BOLOGNA}}
\titlefoot{Centro Brasileiro de Pesquisas F\'{\i}sicas, rua Xavier Sigaud 150,
     BR-22290 Rio de Janeiro, Brazil \\
     \indent~~and Depto. de F\'{\i}sica, Pont. Univ. Cat\'olica,
     C.P. 38071 BR-22453 Rio de Janeiro, Brazil \\
     \indent~~and Inst. de F\'{\i}sica, Univ. Estadual do Rio de Janeiro,
     rua S\~{a}o Francisco Xavier 524, Rio de Janeiro, Brazil
    \label{BRASIL}}
\titlefoot{Comenius University, Faculty of Mathematics and Physics,
     Mlynska Dolina, SK-84215 Bratislava, Slovakia
    \label{BRATISLAVA}}
\titlefoot{Coll\`ege de France, Lab. de Physique Corpusculaire, IN2P3-CNRS,
     FR-75231 Paris Cedex 05, France
    \label{CDF}}
\titlefoot{CERN, CH-1211 Geneva 23, Switzerland
    \label{CERN}}
\titlefoot{Institut de Recherches Subatomiques, IN2P3 - CNRS/ULP - BP20,
     FR-67037 Strasbourg Cedex, France
    \label{CRN}}
\titlefoot{Now at DESY-Zeuthen, Platanenallee 6, D-15735 Zeuthen, Germany
    \label{DESY}}
\titlefoot{Institute of Nuclear Physics, N.C.S.R. Demokritos,
     P.O. Box 60228, GR-15310 Athens, Greece
    \label{DEMOKRITOS}}
\titlefoot{FZU, Inst. of Phys. of the C.A.S. High Energy Physics Division,
     Na Slovance 2, CZ-180 40, Praha 8, Czech Republic
    \label{FZU}}
\titlefoot{Currently at DPNC,
     University of Geneva,
     Quai Ernest-Ansermet 24, CH-1211, Geneva, Switzerland
    \label{GENEVA}}
\titlefoot{Dipartimento di Fisica, Universit\`a di Genova and INFN,
     Via Dodecaneso 33, IT-16146 Genova, Italy
    \label{GENOVA}}
\titlefoot{Institut des Sciences Nucl\'eaires, IN2P3-CNRS, Universit\'e
     de Grenoble 1, FR-38026 Grenoble Cedex, France
    \label{GRENOBLE}}
\titlefoot{Helsinki Institute of Physics, HIP,
     P.O. Box 9, FI-00014 Helsinki, Finland
    \label{HELSINKI}}
\titlefoot{Joint Institute for Nuclear Research, Dubna, Head Post
     Office, P.O. Box 79, RU-101 000 Moscow, Russian Federation
    \label{JINR}}
\titlefoot{Institut f\"ur Experimentelle Kernphysik,
     Universit\"at Karlsruhe, Postfach 6980, DE-76128 Karlsruhe,
     Germany
    \label{KARLSRUHE}}
\titlefoot{Institute of Nuclear Physics and University of Mining and Metalurgy,
     Ul. Kawiory 26a, PL-30055 Krakow, Poland
    \label{KRAKOW}}
\titlefoot{Universit\'e de Paris-Sud, Lab. de l'Acc\'el\'erateur
     Lin\'eaire, IN2P3-CNRS, B\^{a}t. 200, FR-91405 Orsay Cedex, France
    \label{LAL}}
\titlefoot{School of Physics and Chemistry, University of Lancaster,
     Lancaster LA1 4YB, UK
    \label{LANCASTER}}
\titlefoot{LIP, IST, FCUL - Av. Elias Garcia, 14-$1^{o}$,
     PT-1000 Lisboa Codex, Portugal
    \label{LIP}}
\titlefoot{Department of Physics, University of Liverpool, P.O.
     Box 147, Liverpool L69 3BX, UK
    \label{LIVERPOOL}}
\titlefoot{LPNHE, IN2P3-CNRS, Univ.~Paris VI et VII, Tour 33 (RdC),
     4 place Jussieu, FR-75252 Paris Cedex 05, France
    \label{LPNHE}}
\titlefoot{Department of Physics, University of Lund,
     S\"olvegatan 14, SE-223 63 Lund, Sweden
    \label{LUND}}
\titlefoot{Universit\'e Claude Bernard de Lyon, IPNL, IN2P3-CNRS,
     FR-69622 Villeurbanne Cedex, France
    \label{LYON}}
\titlefoot{Univ. d'Aix - Marseille II - CPP, IN2P3-CNRS,
     FR-13288 Marseille Cedex 09, France
    \label{MARSEILLE}}
\titlefoot{Dipartimento di Fisica, Universit\`a di Milano and INFN-MILANO,
     Via Celoria 16, IT-20133 Milan, Italy
    \label{MILANO}}
\titlefoot{Dipartimento di Fisica, Univ. di Milano-Bicocca and
     INFN-MILANO, Piazza delle Scienze 2, IT-20126 Milan, Italy
    \label{MILANO2}}
\titlefoot{IPNP of MFF, Charles Univ., Areal MFF,
     V Holesovickach 2, CZ-180 00, Praha 8, Czech Republic
    \label{NC}}
\titlefoot{NIKHEF, Postbus 41882, NL-1009 DB
     Amsterdam, The Netherlands
    \label{NIKHEF}}
\titlefoot{National Technical University, Physics Department,
     Zografou Campus, GR-15773 Athens, Greece
    \label{NTU-ATHENS}}
\titlefoot{Physics Department, University of Oslo, Blindern,
     NO-1000 Oslo 3, Norway
    \label{OSLO}}
\titlefoot{Dpto. Fisica, Univ. Oviedo, Avda. Calvo Sotelo
     s/n, ES-33007 Oviedo, Spain
    \label{OVIEDO}}
\titlefoot{Department of Physics, University of Oxford,
     Keble Road, Oxford OX1 3RH, UK
    \label{OXFORD}}
\titlefoot{Dipartimento di Fisica, Universit\`a di Padova and
     INFN, Via Marzolo 8, IT-35131 Padua, Italy
    \label{PADOVA}}
\titlefoot{Rutherford Appleton Laboratory, Chilton, Didcot
     OX11 OQX, UK
    \label{RAL}}
\titlefoot{Dipartimento di Fisica, Universit\`a di Roma II and
     INFN, Tor Vergata, IT-00173 Rome, Italy
    \label{ROMA2}}
\titlefoot{Dipartimento di Fisica, Universit\`a di Roma III and
     INFN, Via della Vasca Navale 84, IT-00146 Rome, Italy
    \label{ROMA3}}
\titlefoot{DAPNIA/Service de Physique des Particules,
     CEA-Saclay, FR-91191 Gif-sur-Yvette Cedex, France
    \label{SACLAY}}
\titlefoot{Instituto de Fisica de Cantabria (CSIC-UC), Avda.
     los Castros s/n, ES-39006 Santander, Spain
    \label{SANTANDER}}
\titlefoot{Dipartimento di Fisica, Universit\`a degli Studi di Roma
     La Sapienza, Piazzale Aldo Moro 2, IT-00185 Rome, Italy
    \label{SAPIENZA}}
\titlefoot{Inst. for High Energy Physics, Serpukov
     P.O. Box 35, Protvino, (Moscow Region), Russian Federation
    \label{SERPUKHOV}}
\titlefoot{J. Stefan Institute, Jamova 39, SI-1000 Ljubljana, Slovenia
     and Laboratory for Astroparticle Physics,\\
     \indent~~Nova Gorica Polytechnic, Kostanjeviska 16a, SI-5000 Nova Gorica, Slovenia, \\
     \indent~~and Department of Physics, University of Ljubljana,
     SI-1000 Ljubljana, Slovenia
    \label{SLOVENIJA}}
\titlefoot{Fysikum, Stockholm University,
     Box 6730, SE-113 85 Stockholm, Sweden
    \label{STOCKHOLM}}
\titlefoot{Dipartimento di Fisica Sperimentale, Universit\`a di
     Torino and INFN, Via P. Giuria 1, IT-10125 Turin, Italy
    \label{TORINO}}
\titlefoot{Dipartimento di Fisica, Universit\`a di Trieste and
     INFN, Via A. Valerio 2, IT-34127 Trieste, Italy \\
     \indent~~and Istituto di Fisica, Universit\`a di Udine,
     IT-33100 Udine, Italy
    \label{TU}}
\titlefoot{Univ. Federal do Rio de Janeiro, C.P. 68528
     Cidade Univ., Ilha do Fund\~ao
     BR-21945-970 Rio de Janeiro, Brazil
    \label{UFRJ}}
\titlefoot{Department of Radiation Sciences, University of
     Uppsala, P.O. Box 535, SE-751 21 Uppsala, Sweden
    \label{UPPSALA}}
\titlefoot{IFIC, Valencia-CSIC, and D.F.A.M.N., U. de Valencia,
     Avda. Dr. Moliner 50, ES-46100 Burjassot (Valencia), Spain
    \label{VALENCIA}}
\titlefoot{Institut f\"ur Hochenergiephysik, \"Osterr. Akad.
     d. Wissensch., Nikolsdorfergasse 18, AT-1050 Vienna, Austria
    \label{VIENNA}}
\titlefoot{Inst. Nuclear Studies and University of Warsaw, Ul.
     Hoza 69, PL-00681 Warsaw, Poland
    \label{WARSZAWA}}
\titlefoot{Fachbereich Physik, University of Wuppertal, Postfach
     100 127, DE-42097 Wuppertal, Germany
    \label{WUPPERTAL}}
\clearpage
\end{titlepage}
%%%%%%%%%%%%%%%%%%%%%%%%%
%
% Change for the document body
%%\pagestyle{heading} % for page numbering
\pagenumbering{arabic} % page numbering in number
\renewcommand{\thefootnote}{\fnsymbol{footnote}} % symbolic footnote marks
\setcounter{footnote}{1} %
\large
%\linenumbers %%%CD
\newcommand{\stau} {$\tilde{\tau}$}
\newcommand{\stuno} {$\tilde{\tau}_1$}
\newcommand{\staur} {$\tilde{\tau}_R$}
\newcommand{\staul} {$\tilde{\tau}_L$}
\newcommand{\selr} {$\tilde{e}_R$}
\newcommand{\smur} {$\tilde{\mu}_R$}
\newcommand{\slep} {$\tilde{l}$}
\newcommand{\slepr} {$\tilde{l}_R$}
\newcommand{\nuno} {$\tilde{\chi}^0_1$}
\newcommand{\chino} {$\tilde{\chi}^\pm_1$}
\newcommand{\grav} {$\tilde{G}$}
\newcommand{\mgrav} {$m_{\tilde{G}}$}
\newcommand{\ra} {\rightarrow}
\newcommand{\eeto} {\mbox{$ {\mathrm e}^+ {\mathrm e}^-\! \ra\ $}}
\newcommand{\ee} {\mbox{$ {\mathrm e}^+ {\mathrm e}^-$}}
\newcommand{\qqbar} {$q\bar{q}$}
\newcommand{\Wp} {\mbox{$ {\mathrm W}^+$}}
\newcommand{\Wm} {\mbox{$ {\mathrm W}^-$}}
\newcommand{\Zn} {\mbox{$ {\mathrm Z}$}}
\newcommand{\Wev} {\mbox{$ {\mathrm{W e}} \nu_{\mathrm e}$}}
\newcommand{\Zvv} {\mbox{$ \Zn \nu \bar{\nu}$}}
\newcommand{\Zee} {\mbox{$ \Zn \ee$}}
\newcommand{\GeV} {~\mbox{${\mathrm{GeV}}$}}
%KH USUALLY HAVE "C" IN ITALICS, SINCE IT'S A PHYSICAL QUANTITY...
%KH \newcommand{\GeVcc} {~\mbox{${\mathrm{GeV}}/{\mathrm{c}}^2$}}
%KH \newcommand{\eVcc} {~\mbox{${\mathrm{eV}}/{\mathrm{c}}^2$}}
%KH \newcommand{\MeVc} {~\mbox{$ {\mathrm{MeV}}/ {\mathrm{c}} $}}
%KH \newcommand{\MeVcc} {~\mbox{$ {\mathrm{MeV}}/{\mathrm{c}}^2 $}}
\newcommand{\GeVcc} {~\mbox{${\mathrm{GeV}}/c^2$}}
\newcommand{\eVcc} {~\mbox{${\mathrm{eV}}/c^2$}}
\newcommand{\MeVc} {~\mbox{$ {\mathrm{MeV}}/c $}}
\newcommand{\GeVc} {~\mbox{$ {\mathrm{GeV}}/c $}}
\newcommand{\MeVcc} {~\mbox{$ {\mathrm{MeV}}/c^2 $}}
\newcommand{\TeV} {~\mbox{$ {\mathrm{TeV}} $}}
%KH ADDED keV/c^2 (LOWER CASE "k" FOR "KILO" IS SI STANDARD)
\newcommand{\keVcc} {~\mbox{$ {\mathrm{keV}}/c^2 $}}
\newcommand{\etal} {\mbox{\it et al.}}
\def\NPB#1#2#3{{\rm Nucl.~Phys.} {\bf{B#1}} (19#2) #3}
\def\PLB#1#2#3{{\rm Phys.~Lett.} {\bf{B#1}} (19#2) #3}
\def\PRD#1#2#3{{\rm Phys.~Rev.} {\bf{D#1}} (19#2) #3}
\def\PRL#1#2#3{{\rm Phys.~Rev.~Lett.} {\bf{#1}} (19#2) #3}
\def\ZPC#1#2#3{{\rm Z.~Phys.} {\bf C#1} (19#2) #3}
\def\PTP#1#2#3{{\rm Prog.~Theor.~Phys.} {\bf#1} (19#2) #3}
\def\MPL#1#2#3{{\rm Mod.~Phys.~Lett.} {\bf#1} (19#2) #3}
\def\PR#1#2#3{{\rm Phys.~Rep.} {\bf#1} (19#2) #3}
\def\RMP#1#2#3{{\rm Rev.~Mod.~Phys.} {\bf#1} (19#2) #3}
\def\HPA#1#2#3{{\rm Helv.~Phys.~Acta} {\bf#1} (19#2) #3}
\def\NIMA#1#2#3{{\rm Nucl.~Instr.~and~Meth.} {\bf#1} (19#2) #3}
\def\CPC#1#2#3{{\rm Comp.~Phys.~Comm.} {\bf#1} (19#2) #3}

\section{Introduction}
\label{sec:intro}
In 1999, the centre-of-mass energies reached by LEP ranged from 192 GeV to
202 GeV, and the DELPHI experiment collected an integrated lu\-mi\-no\-si\-ty 
of \mbox{228.2 pb$^{-1}$}. These data were analysed to update the searches
for sleptons, neutralinos and charginos in the context of Gauge Mediated
Supersymmetry Breaking (GMSB) models~\cite{Dine1,Dine2} already performed
at lower energies. 

In these models
the gravitino, $\tilde{G}$, is the lightest supersymmetric particle
(LSP) and the next-to-lightest supersymmetric particle (NLSP) can be either the
neutralino, \nuno, or the sleptons,
\slep~~\cite{Bagger,Dutta,Cheung,francesca,giudice}.
The data were analysed under the assumption that the NLSP is a
slepton. Depending on the magnitude of the mixing
between the left and right gauge eigenstates, \staur~and
\staul, there are two possible 
scenarios. If the mixing is large\footnote{In GMSB models large mixing 
occurs generally in regions of $\tan\beta\geq 1$0 (the ratio of the vacuum
expectation values of the two Higgs doublets) or $|\mu|>50$0 \GeVcc
~($\mu$ is the Higgs mass parameter).}, 
\stuno\ (the lighter mass eigenstate)
% which is a combination of the two gauge eigenstates) 
is the NLSP. However, if the mixing 
is negligible, \stuno\ is mainly right-handed~\cite{bartl}
and almost mass degenerate with the other sleptons. In this case, the
\selr\ and \smur\ three 
body decay (\slep $\rightarrow$ \stuno $\tau l$ with
\stuno $\rightarrow~\tau$ \grav), is very suppressed, and \selr~and
\smur~decay directly into $l$\grav. This scenario is called sleptons co-NLSP.

Due to the coupling of the NLSP to \grav, the mean decay length, $L$, of the NLSP
can range from micrometres to metres depending on the mass of the
gravitino~\cite{martin}
(\mgrav):
\begin{equation}
L = 1.76 \times 10^{-3} \sqrt{\left (\frac{E_{\tilde{l}}}{m_{\tilde{l}}}
\right )^2-1}
\left ( \frac{m_{\tilde{l}}}{100 \, {\rm GeV/c}^2} \right )^{-5}
\left ( \frac{m_{\tilde{G}}}{1\, \rm eV/c^2}\right )^{2} \;\; {\mathrm cm}
\label{life}
\end{equation}

For example, for $m_{\tilde{G}}~\lesssim~250$\eVcc, or equivalently, for a
SUSY breaking scale of $\sqrt{F} \lesssim$ 1000\TeV~(since both parameters
are related~\cite{gravitino}),  the decay of a NLSP with mass greater than
for example 60\GeVcc~can take place
within the detector. This range of $\sqrt{F}$ is in fact consistent with
astrophysical and cosmological considerations \cite{Dinopoulos0,wagner}.

In this work the results of the searches reported
in~\cite{nuestro_papel_nuevo} are updated and the search for
charginos is extended to the sleptons co-NLSP scenario.
Moreover, the update of the search for heavy stable charged
particles presented in~\cite{heavyparticles} is also
performed.  Heavy stable charged particles are predicted not only in GMSB
  models but also in Minimal Supersymmetric Standard Models (MSSM) with a very
  small amount of R-parity violation, or with R-parity conservation 
 if the mass difference between the LSP and the NLSP becomes very small. In
 these models the LSP can be a 
  charged slepton or a squark and decay with a long lifetime into Standard
  Model particles~\cite{Dreiner}. Therefore, updated lower mass limits on
  heavy stable charged particles, under the assumption that 
  the LSP is a charged
  slepton, will be provided in this letter within both models, GMSB and MSSM.

The first search  looks for the production of \nuno~pairs in the \stuno~NLSP
scenario 
\begin{equation}
e^+e^- \to \tilde{\chi}^0_1 \tilde{\chi}^0_1 \to 
\tilde{\tau}^+_1 \tau^- \tilde{\tau}^+_1 \tau^-  \to \tau^+ \tilde{G} \tau^-
\tau^+ \tilde{G} \tau^- 
\label{eq:nunostuno}
\end{equation}
and in the sleptons co-NLSP scenario 
\begin{equation}
e^+e^- \to \tilde{\chi}^0_1
\tilde{\chi}^0_1 \to  
\tilde{l}^+_R l^- \tilde{l}'^+_R l'^-  \to l^+ \tilde{G} l^- l'^+ \tilde{G}
l'^- 
\label{eq:nunoslept}
\end{equation}
with $l$ = $e$, $\mu$, $\tau$ and
BR($\tilde{\chi}^0_1 \to \tilde{l} l) = 1/3$ for each leptonic flavour.
In the former 
case,  neutralino pair production would mainly lead to 
a final state with four tau leptons and two gravitinos, while 
in the case of a co-NLSP scenario, the final signature would contain two
pairs of leptons with possibly different flavour and two gravitinos.

The second search concerns $\tilde{l}$ pair production followed
by the decay of each slepton into a lepton and a gravitino:
\begin{equation}
e^+e^- \to \tilde{l}^+\tilde{l}^- \to l^+ \tilde{G} l^- \tilde{G} 
\label{eq:kink}
\end{equation}
This search has been performed within the
\stuno~NLSP scenario ($\tilde{l}$ = $\tilde{\tau}_1$) 
and the sleptons co-NLSP scenario ($\tilde{l}$ = $\tilde{l}_R$).
The signature of these events will depend on the mean decay length of the
NLSP, or 
equivalently, on the gravitino mass. Therefore, if the decay length is
too short (1\eVcc~$\lesssim m_{\tilde{G}} \lesssim$ 10\eVcc) 
to allow the reconstruction of the slepton,
only the corresponding lepton or its decay products will be seen in
the detector, and the search will then be based on the track impact parameter.
If the slepton decays inside the tracking devices 
(10\eVcc~$\lesssim m_{\tilde{G}} \lesssim$ 1000\eVcc), the signature 
will be at least one track of a charged particle with a kink or a decay vertex.
However, for very heavy gravitinos ($m_{\tilde{G}} \gtrsim$ 1000\eVcc), the
decay length is large and the slepton decays outside the detector. The pair
production of such long-lived or stable particles yields a characteristic
signature with typically two back-to-back charged heavy objects in the
detector.  Finally, for very light gravitino masses 
($m_{\tilde{G}} \lesssim$ 1\eVcc), the decay takes place in the primary vertex 
and the results from the search for sleptons
in gravity mediated (MSUGRA) models can be applied~\cite{slep189}.

In the parameter space where the sleptons are the NLSP, there are specific
regions where
the chargino is light enough to be produced at LEP~\cite{Cheung}. Therefore,
the third search looks for the pair production of lightest charginos
in the \stuno~NLSP scenario 
\begin{equation}
e^+e^- \to \tilde{\chi}^+_1 \tilde{\chi}^-_1 \to 
\tilde{\tau}^+_1 \nu_{\tau} \tilde{\tau}^-_1 \overline{\nu}_{\tau}  \to
\tau^+ \tilde{G} \nu_{\tau}
\tau^- \tilde{G} \overline{\nu}_{\tau}  
\label{eq:chargistuno}
\end{equation}
and sleptons co-NLSP scenario 
\begin{equation}
e^+e^- \to \tilde{\chi}^+_1 \tilde{\chi}^-_1 \to 
\tilde{l}^+_R \nu_l \tilde{l'}^-_R \overline{\nu}_{l'}  \to l^+ \tilde{G} \nu_l
l'^- \tilde{G} \overline{\nu}_{l'}. 
\label{eq:chargislept}
\end{equation}
The analysis is divided into four
topologies according to the mean lifetime 
of the slepton as explained in the previous paragraph: 
two acoplanar leptons with respect to the beam pipe
with missing energy (MSUGRA models), at least one track with large impact
parameter or a  
kink, or at least one track corresponding to a very massive stable charged 
particle.

The data samples are described in 
section~\ref{experimentalprocedure}. The efficiencies of the different
selection criteria and the number of events selected in data and in the
Standard Model background are reported in~\ref{dataselection}. Finally, the
results are presented in section~\ref{sec:resultados}. 
%
%------------------------------------------------------------
% EVENT SAMPLE
%------------------------------------------------------------
%
\section{Data sample and event generators}
\label{experimentalprocedure}

All searches are based on 
data collected with the DELPHI detector
 during 1999 at 
centre-of-mass energies around 192, 196, 200 and 202 GeV.
The total integrated lu\-mi\-no\-si\-ty was 228.2~pb$^{-1}$.  
A detailed
description of the DELPHI detector can be found in \cite{detector} and the
detector performance in \cite{performance}.

To evaluate the signal efficiencies and background contamination,
events were ge\-ne\-ra\-ted using different programs, all
relying on {\tt JETSET} 7.4 \cite{JETSET}, tuned
to LEP~1 data \cite{TUNE} for quark fragmentation.

The program {\tt SUSYGEN} \cite{SUSYGEN} was used to generate
neutralino pair events and their subsequent decay products. 
In order to compute detection efficiencies, a
total of 90000 events were generated 
with masses \mbox{67\GeVcc$\leq m_{\tilde{\tau}_1}+2$\GeVcc~$\leq
m_{\tilde{\chi}^0_1} \leq \sqrt{s}/2$} and at the 
four centre-of-mass energies. 

Slepton pair samples of 99000 and 76500 events at 196 GeV and 202 GeV
centre-of-mass energies 
respectively were
 produced with {\tt PYTHIA} 5.7\footnote{Another two samples of 1000 events
   with 
  $m_{\tilde{\tau}}$ = 60 \GeVcc~and mean decay lengths of 5 and 50~cm, were
  generated using {\tt SUSYGEN} to cross check with the {\tt PYTHIA} results. 
The same efficiencies were found within a $\pm$2\% difference.} \cite{JETSET} with
staus having a mean decay length from  0.25 to 200 cm and masses
from 40 to 100\GeVcc. 
Other samples of \stau\ pairs were produced with {\tt SUSYGEN} for the 
small impact parameter search with $m_{\tilde{\tau}}$\ from 40\GeVcc~to 
100\GeVcc.

For the search for heavy stable charged particles, signal efficiencies were
estimated on the basis of about 50000 simulated events.
Pair produced heavy smuons were generated
at energies between 192~GeV and 202~GeV with {\tt SUSYGEN}, and passed
through the detector simulation as heavy muons. The efficiencies were estimated
for masses between 10\GeVcc~and 97.5\GeVcc. 1000 events were generated per mass point.

{\tt SUSYGEN} was also used to generate
 \chino\ pair production samples and their decays at 192~GeV and 202~GeV
centre-of-mass energies. 
In order to compute detection efficiencies, a
total of 45 samples with 500 events each were generated 
with \mgrav~at 1, 100 and 1000\eVcc ,
\mbox{$ m_{\tilde{\tau}_1}+ 0.3$\GeVcc~$\leq m_{\tilde{\chi}^+_1} \leq 
\sqrt{s}/2$} and
\mbox{$m_{\tilde{\tau}_1}\geq 65$\GeVcc}. 
Samples with \mbox{smaller} \mbox{$\Delta m = 
m_{\tilde{\chi}^+_1} -m_{\tilde{\tau}_1}$} were not generated because 
in this region the \chino\ does not decay mainly to 
$\tilde{\tau}_1$ and $\nu_{\tau}$ but into W and $\tilde{G}$.
%The different background samples and event selections are described in 
%references~\cite{chargino_msugra,slep_200} and~\cite{heavyparticles} 
%for $m_{\tilde{G}} = $1 and 
%1000\eVcc\ respectively. For the case of $m_{\tilde{G}} = 100$\eVcc , the
%analysis is the same as  the search for sleptons in this paper
%and consequently the same samples of simulated background events were used.  

The background process \eeto\qqbar ($n\gamma$) was generated with
{\tt PYTHIA 6.125}, while {\tt KORALZ 4.2} \cite{KORALZ} was used
for $\mu^+\mu^-(\gamma)$ and $\tau^+\tau^-(\gamma)$.
The generator {\tt BHWIDE}~\cite{BHWIDE} was used for \eeto\ee\ events.

Processes leading to four-fermion final states
were generated using {\tt EXCALIBUR 1.08}~\cite{EXCALIBUR} and {\tt
  GRC4F}~\cite{GRC4F}. 

Two-photon interactions leading to hadronic final states
%leading to hadronic and leptonic final states
were generated using {\tt TWOGAM}~\cite{TWOGAM}, including the VDM, QPM and
QCD components.
The generators
of Berends, Daverveldt and Kleiss~\cite{BDK} were used for the leptonic
final states.

The cosmic radiation background was studied using the data collected
before the beginning of the 1998 LEP run.

The generated signal and background events were passed through the
detailed simulation~\cite{performance}
%~\cite{delsim} 
of the DELPHI detector 
%\cite{detector} 
and then processed
with the same reconstruction and analysis programs used for real 
data.

%
%------------------------------------------------------------
% EVENT SELECTION
%------------------------------------------------------------
%
\section{Data selection}
\label{dataselection}

\subsection{Neutralino pair production}
\label{neutra}

The selection criteria used in the search for neutralino pair production in
the \stuno~NLSP scenario  and in the sleptons co-NLSP scenario, 
were described in detail
in~\cite{nuestro_papel_nuevo,nuestro_papel}. 
The main two differences between these two cases are
that the mean 
number of neutrinos carrying away undetected energy and momentum and the 
number of charged particles per event are considerably bigger for the
\stuno~NLSP  scenario.

After applying the selection criteria to the search for these topologies, 
six events passed the search for neutralino pair
production in  
the \stuno~NLSP scenario, with 3.36$\pm$0.98 Standard Model (SM)
background events expected. Four events passed the search for neutralino
pair  production in the sleptons co-NLSP scenario, with 4.39$\pm$0.51 SM
background events expected.
Efficiencies between 20\% and 44\% were obtained for the signal events.

\subsection{Slepton pair production}
\label{stau}
This section describes the update of the search for slepton pair 
production as a function of the mean decay length. The details of the 
selection criteria used to search for the topologies obtained when the 
NLSP decays inside the detector volume were described
in~\cite{nuestro_papel_nuevo,nuestro_papel,ref:flying}. Likewise, 
the selection criteria used to search for heavy stable charged particles
were described in detail in~\cite{heavyparticles,heavyone}.
The efficiencies were derived for different \slep~masses and
decay lengths by applying the same selection to
the simulated signal events.

\subsubsection{Search for secondary vertices}
\label{searchkinks}
\hspace{\parindent}

This analysis exploits a feature of the
$\tilde{l} \rightarrow l \tilde{G}$ topology when the slepton decays inside
the tracking devices,
%in the case of intermediate gravitino masses 
%(i.e. 0.5\eVcc\ $\lesssim m_{\tilde{G}}\lesssim$\ 200\eVcc\ as dictated by 
%equation~\ref{life}), 
namely, one or two tracks originating from the interaction point and at least one
of them with a secondary vertex or a kink. 
After applying the selection criteria to search for this topology,
two events in real data were found to satisfy all the requirements, 
while 0.79$^{+0.28}_{-0.12}$\ were expected from SM backgrounds. 
One event was compatible with a $\gamma \gamma \rightarrow 
\tau^+\tau^-$ with a hadronic interaction in the Inner detector. The other one
was compatible with a $e^+e^-~\rightarrow~\tau^+\tau^-$ event where one of
the electrons (a decay product of the $\tau$), after
radiating a photon, was reconstructed as two independent tracks.   

The secondary vertex reconstruction procedure was sensitive to radial decay
lengths, R,  
between 20~cm and 90~cm. 
%Within this region the search for events with at
%least one secondary vertex detected had 
%an efficiency of 54.0$\pm$2.0\%. 
The  Vertex detector and the Inner detector  
were needed to reconstruct the
$\tilde{\tau}$ and the Time Projection Chamber
to reconstruct the decay products. 
The shape of the efficiency distribution was essentially flat as a 
function of R decreasing when
the \stau\ decayed near the outer surface of the Time Projection Chamber. The
decrease was due to inefficiencies  
in the reconstruction of the tracks coming from the decay products of 
the $\tau$. The search for events with secondary vertices had 
an efficiency of (52.8$\pm$2.0)\% for \stau\ with masses between 40 and 
100\GeVcc, with a mean decay length of 50~cm. 
%The efficiencies decreased 
%near the kinematical limit due to a small boost that allowed for larger 
%angles to appear between the \stau\ and the decay products of the $\tau$.
%For \stau\ masses below 40\GeVcc, the efficiency decreased gradually 
%due to the cut that rejects segmented tracks. This happened 
%because the resulting big boost causes the angle between \stau\ and 
%$\tau$\ decay products to be very small. 

The same selection criteria were applied to smuons and
selectrons. The  
efficiency was (56.5$\pm$2.0)\% for $m_{\tilde{\mu}_R}$  between 40
to 100\GeVcc, and for selectrons it was (38.1$\pm$2.0)\% in the same range of
masses. The efficiency for selectrons was lower than for staus or smuons
due to an upper cut  on total electromagnetic energy at the preselection level.
 
\subsubsection{Large impact parameter search}
\label{searchimpact}
\hspace{\parindent}

To investigate the region of lower gravitino masses 
the previous search was extended to the case of sleptons with mean decay length
between 0.25 cm and approximately 10~cm. In this case the \slep~is not
reconstructed and only the $l$ (or the decay products in the case of \stau)
is detected. 
The impact parameter search was only applied to those events accepted by the
same general requirements as in the search for secondary vertices, and not 
selected by the secondary vertex analysis. 

The maximum efficiency was (31.0$\pm$2.0)\% corresponding to a mean decay
length of 2.5~cm. The efficiency decreased sharply for lower decay lengths
due to the requirement on mi\-ni\-mum impact parameter. 
For longer decay lengths, 
the appearance of reconstructed $\tilde{\tau}$ in combination with
the cut on the maximum number of charged particles in the event caused the
efficiency to decrease smoothly. This decrease is compensated by a rising
efficiency in the search for secondary vertices.
For masses above 40\GeVcc\ no dependence on the  $\tilde{\tau}$ mass was 
found far from the kinematic limit. 
%However for lower masses the efficiency
%decreased and it was almost zero for a 5\GeVcc\ \stuno.

The same selection was applied to smuons and selectrons. For smuons
the efficiency increased to (60.0$\pm$2.0)\% for a mean decay 
length of 2.5 cm and masses over 
40\GeVcc\ since the smuon always has a one-prong decay.
For selectrons the efficiency was (36.3$\pm$2.0)\% for the same mean decay
length and range of masses.  

Trigger efficiencies were studied simulating the DELPHI trigger
response to the events selected by the vertex search and by the large impact
parameter analysis, and were found to be around 99\%.

Two events in the real data sample were selected, while
1.77$^{+0.25}_{-0.21}$\ were expected from SM backgrounds.
Both events are compatible with $e^+e^-~\rightarrow~\tau^+\tau^-$ events
where one of 
the electrons (a decay product of the $\tau$), after 
radiating a photon, was reconstructed only by the Inner detector and Time
Projection Chamber detectors giving
a very large impact parameter track. 
\noindent

%\newpage
\subsubsection{Small impact parameter search}
\label{smallimpact}
\hspace{\parindent}

The large impact parameter search can be extended further to 
mean decay lengths below 0.1 cm. 
Here only the main points of the analysis and some 
changes with respect to previous ones are recalled. In low multiplicity events
two hemispheres were defined using the thrust axis. The highest momentum,
good quality ($\Delta$p/p $< 50\%$), particle tracks in each hemisphere were 
labelled leading tracks. The impact parameters from the beam spot, 
$b_1$ and $b_2$, of the 
leading tracks in the R$\phi$ plane were used to discriminate against 
SM backgrounds.  
The same selection criteria described in 
references~\cite{nuestro_papel_nuevo,nuestro_papel} were applied. However,
some  extra selection was added in order to 
reduce the background from detector noise or failure.

In order to preserve the efficiency in the region of decay length $\gtrsim$ 10
cm, where the $\tilde \tau$ can be observed as a particle coming from the
primary 
vertex and badly measured owing to its limited length, further requirements 
on the track quality were applied only to the leading track with the larger
impact parameter. This particle was required to have a relative momentum error 
$< 30\%$ and the track to be measured at least either in 
the Time Projection Chamber or in all of the other three track
detectors in the barrel (Vertex detector, Inner detector and Outer detector).

The efficiency of the search did not show any significant dependence 
on the $\tilde{\tau}$ mass for masses over 40 GeV/$c^2$ and it could be 
parameterized as a function of  
the $\tilde{\tau}$ decay length in the laboratory system. 
The maximum efficiency was $\sim$ 38\% for a mean decay length of 
$\sim$ 2~cm; the efficiency dropped 
at small decay lengths ($\sim$ 10\% at 0.6~mm).

The same selection criteria were used to search for smuons
as reported in~\cite{nuestro_papel_nuevo}.
The maximum efficiency reached for the smuon search was $\sim$ 43\% at
$\sim$ 2~cm mean decay length. To search for
selectrons, in order to increase efficiency, the cut 
\mbox{$(E_1+E_2)<0.7~E_{beam}$}\ 
(where $E_1$, $E_2$ are the electromagnetic energy deposits associated to the
leading tracks) was not applied. The Bhabha events that survived the
selection, when the previous rejection cut was not applied, were those where at
least one of the electrons un\-der\-went a secondary interaction, thus acquiring
a large impact parameter. However, it was found that in these cases the
measured momentum of the electron was smaller than the electromagnetic energy
deposition around the electron track. Therefore, the cut 
 $(E_1/p_1+E_2/p_2)<2.2$ was used for the selectron search.
The maximum efficiency reached in the 
selectron search was $\sim$ 35\% at  $\sim$ 2~cm mean decay length.

Requiring $\sqrt{b_1^2 + b_2^2}~>$ 600 $\mu$m, 
the number of events selected in the data was 5 in the $\tilde{\tau}$ and
$\tilde{\mu}$ search, and 4 in the $\tilde{e}$ search, while
5.05$\pm 0.39$ events were expected 
from the SM background in both searches. 
%Figure~\ref{fig:ip-data-mc} shows the $\sqrt{b_1^2 + b_2^2}$ distribution for
%data (dots) and simulated  
%backgrounds (histogram) after all other cuts. 
All of the selected candidates were compatible with SM 
events. 

\subsubsection{Heavy stable charged particle search}
\label{heavystable}
\hspace{\parindent}
The analysis described in \cite{heavyparticles} has been applied for 
each of the four centre-of-mass energies 192, 196, 200 and 202~GeV. 
A careful run selection ensured that the Ring Imaging CHerenkov (RICH) 
detectors were fully operational because the method used to identify 
heavy stable particles relies on the lack of Cherenkov radiation 
in DELPHI's RICH detectors. The
luminosities analysed after the run selection are shown
in Table~\ref{tab:lumi}.
Signal efficiencies
were estimated from Monte Carlo by simulating heavy sleptons with {\tt SUSYGEN}
and passing them through the detector simulation as heavy muons. The background given in Table ~\ref{tab:lumi} was estimated from data itself by
counting the number of tracks passing the individual selection criteria.
Only events with two or three charged particles were considered. 
Events were selected, if they contained at least one charged particle with:
\begin{enumerate}
\item a momentum above 5~GeV/$c$, a high ionization
loss and no photons in the gas radiator were associated to the particle 
(gas veto) or, 
\item a momentum above 15~GeV/$c$, an ionization loss
 at least 0.3 below the expectation for a proton and surviving the gas veto
 or,  
\item a momentum above 15~GeV/$c$, surviving the gas
and the liquid RICH veto.
\end{enumerate}
An event was also selected if both event hemispheres contained particles with
both either a high ionization loss or a gas veto, or both having a
low ionization loss.

No candidate events were selected in data.
As an example, Figure~\ref{rare202} show the data and the three main
search windows for the search at an energy of 202~GeV.
The expectation for a 90~GeV/$c^2$ mass signal is also shown.
For particle masses below 60~GeV/$c^2$ the signal efficiencies are of the order
of 30\%, and rise with increasing mass to about 78\%.
Then the efficiency
drops when approaching the kinematical limit due to saturation effects,
and it is assumed to be zero at the kinematical limit.

\begin{table}[tbh]
\begin{center}
\begin{tabular}{l|c|c|c|c}
$\sqrt{s}$   & 192 GeV & 196 GeV & 200 GeV & 202 GeV \\
\hline
${\cal L}$ (pb$^{-1}$) & 26.3& 69.7& 87.1& 40.4\\
background & 0.12$\pm$0.04 & 0.18$\pm$0.04 & 0.31$\pm$0.06& 0.1$\pm$0.03 \\
observed   & 0             & 0             & 0            & 0 \\
\end{tabular}
\end{center}
\caption{Luminosities analysed and selected events for each of the
four centre-of-mass energies in the search for heavy stable charged particles.}
\label{tab:lumi}
\end{table}

\subsection{Chargino pair production}
\label{chargino}

The search for the lightest chargino depends on the slepton lifetime, or
equivalently on the gravitino mass as already stated.
For $m_{\tilde{G}} \lesssim$ 1\eVcc, \chino~decays at the vertex and the final
state is two acoplanar leptons with missing energy. In this case the search
for charginos and for sleptons 
in gravity mediated scenarios (MSUGRA) can be
applied. The details of the search for charginos in MSUGRA models can be found in~\cite{chargino_msugra}.
The efficiencies obtained using these analyses were 13-36\%
for the $\tilde{\tau}_1$ NLSP scenario and 15-29\% for the 
sleptons co-NLSP scenario. 
The number of selected events in data was
39, while the background expected from SM was 37.58$^{+2.02}_{-0.90}$ for the 
$\tilde{\tau}_1$ NLSP scenario. For the sleptons co-NLSP scenario the
number of candidates in data was 81, while the number of events
expected from SM was 79.7$\pm$3.9. 
For $m_{\tilde{G}}$ between 1\eVcc~and 1000\eVcc, \chino~has
intermediate mean decay lengths and the final topologies are events with
kinks or large impact parameter tracks. 
The efficiencies obtained with these analyses were 25-56\% in the  
$\tilde{\tau}_1$ NLSP scenario, and 41-56\% in the 
sleptons co-NLSP scenario. 
Four events in real data were found to satisfy all the conditions required in
the search for charginos with intermediate mean decay lengths, while
2.56$^{+0.38}_{-0.24}$\ events were expected  from SM 
backgrounds. These results apply to both scenarios \stuno~NLSP
and  sleptons co-NLSP,
since the same selection criteria were applied to search for 
staus, smuons and selectrons.

Finally, for $m_{\tilde{G}} >$ 1000\eVcc~the event topology is
at least one track corresponding to a very massive stable charged 
particle. Therefore
the search for stable charged
particles was applied. In this case the efficiency
is mainly 
affected by the momentum of the slepton because the method used to identify 
heavy stable particles relies on the lack of Cherenkov radiation 
in DELPHI's RICH detectors as already stated in section~\ref{heavystable}. 
To remove SM backgrounds, low momentum particles 
are removed, thus reducing the efficiency for higher chargino masses, 
especially in the region where the mass difference ($\Delta m$) between the
NLSP and the LSP  is small. Therefore for this analysis
the efficiencies vary from 0\% for \mbox{$\Delta m$ = 300\MeVcc}~to 62\% for
\mbox{$\Delta m$ = 20\GeVcc}.
No candidates
in the data passed the selection cuts, while 0.71$\pm$0.09 events were expected
from background. 

\section{Results and interpretation}
\label{sec:resultados}

Since there was no evidence for a signal above the expected background, the
number of candidates in data and the expected number of background events
were used to set limits at the 95\% confidence level (CL) on the pair
production cross-section and masses of the sparticles searched for.
The model described in reference~\cite{Dutta} was
used to derive limits within the GMSB scenarios. This model  
assumes radiatively broken electroweak symmetry and 
null trilinear couplings at the messenger scale. The 
corresponding pa\-ra\-me\-ter space was scanned as follows:
$1\leq n \leq 4$, $5\ {\rm TeV}\leq\Lambda\leq 90\ {\rm TeV}$, 
$1.1\leq M/\Lambda \leq 10^9$, $1.1\leq \tan\beta\leq 50$, and 
$sign(\mu) = \pm 1 $,
where $n$\ is the number of messenger generations in the model, $\Lambda$\ 
is the ratio between the vacuum expectation values 
of the auxiliary component and the scalar component of the
superfield and $M$\ is the messenger mass scale. The parameters 
tan $\beta$\ and $\mu$\ are defined as for MSUGRA.
The limits presented here are at \mbox{$\sqrt{s}$ = 202 GeV}
after combining the results of the searches at lower centre-of-mass energies
%$\sqrt{s}$ = 130-202 GeV 
with the likelihood ratio method~\cite{Read}.

\subsection{Neutralino pair production}

Limits for neutralino pair production cross-section were derived in 
the \stuno~NLSP and sleptons co-NLSP scenarios for
each ($m_{\tilde{\chi}_1^0}$,$m_{\tilde{l}}$) combination. For the 
\stuno~NLSP case, the combination took into account 
the results from the LEP runs from 1996 (for \mbox{$\sqrt{s}~\ge$ 161 GeV}) to
1999.  
The limits for the production cross-section allowed 
some sectors of the ($m_{\tilde{\chi}^0_1},m_{\tilde{l}}$) space to be
excluded. In order to exclude as much as possible of the mass plane, the
results from two other analyses were taken into account. 
The first is the search for slepton pair production
in the context of MSUGRA models.  
In the case where the MSUGRA $\tilde{\chi}_1^0$\ is 
massless, the kinematics correspond to the case of $\tilde{l}$\ 
decaying into a lepton and a gravitino.
The second is the search for lightest neutralino pair production 
in the region of the mass space where  
$\tilde{\chi}_1^0$\ is the NLSP~\cite{2gamma_189} 
(the region above the diagonal 
line in Figure~\ref{fig:masses}, i.e. $m_{\tilde{\tau}_1} >
m_{\tilde{\chi}^0_1}$).  
Within this zone, the neutralino decays into a gravitino and a photon. 

As an illustration, Figure~\ref{fig:masses}
presents the 95\% CL excluded areas for $m_{\tilde{G}} <$ 1\eVcc~in the 
$m_{\tilde{\chi}_1^0}$ vs $m_{\tilde{\tau}_1}$ plane for the \stuno~NLSP. 
The negative-slope dashed area is excluded by
the analysis searching for neutralino pair production followed by the decay
$\tilde{\chi}^0_1\rightarrow \tilde{G}\gamma$. 
The point-hatched area is excluded by the direct search for slepton pair
production within MSUGRA scenarios.
%%%%%%%%%%%%%%%%%%%%%%%%%%%%%%

\subsection{Slepton pair production}

The results of the search for slepton pair production are presented in the 
($m_{\tilde{G}}$,$m_{\tilde{l}}$) plane combining the two impact
parameter searches, the secondary vertex analysis and the stable heavy lepton
search, and using all DELPHI data from 130 GeV to 202 GeV
centre-of-mass energies.
 
The $\tilde{\tau}_1$ pair production cross-section depends on the mixing in
the stau sector. Therefore, in order to put limits on the $\tilde{\tau}_1$
mass the mixing angle had to be fixed. The results presented here correspond
to the case when there is no mixing between the 
\staur~and \staul, thus $\tilde{\tau}_1$ is a pure right-handed state
(Figure~\ref{fig:excl_202}-a). In the case which
corresponds to a mixing angle 
which gives the minimum $\tilde{\tau}_1$ pair production cross-section and
at the same time maintains $m_{\tilde{\tau}_1}^2 >$ 0, the given limit was
reduced by $\sim$~1~GeV. 
Therefore, within the \stuno~NLSP scenario, the
impact parameter and secondary vertex analyses extended the limit  
$m_{\tilde{\tau}_R} >$ 75\GeVcc~for {$m_{\tilde{G}} \lesssim$
  1\eVcc, set by 
MSUGRA searches~\cite{slep189}, up to \mbox{$m_{\tilde{G}}$ = 400\eVcc},
reaching the maximum excluded 
value of
\mbox{$m_{\tilde{\tau}_R}$= 92\GeVcc}~for \mbox{$m_{\tilde{G}}$= 200\eVcc}. 
For 
\mbox{$m_{\tilde{G}} >$ 250\eVcc}~the best lower mass limit was set by the
stable heavy lepton search. 
%excluded $\tilde{\tau}_R$ 
%with a mass below 88\GeVcc\ for gravitino masses between 20 and
%300\eVcc~at 95\%~CL. For \mgrav\ 
%below a few \eVcc, $m_{\tilde{\tau}_R} < 7$6\GeVcc\ was excluded
%by the search for \staur\ in gravity mediated models~\cite{slep_200}.
%For $m_{\tilde{G}}$ larger than 1000\eVcc\
%the lower limit was 93\GeVcc~obtained from the
%stable heavy lepton search~\cite{heavyparticles-conf}.

Within the sleptons co-NLSP scenario, the cross-section limits were used to
derive lower limits for \smur\ (Figure~\ref{fig:excl_202}-b) and \selr\
(Figure~\ref{fig:excl_202}-c) masses at 
95\%~CL. The $\tilde{\mu}_R$ pair production
cross-section is model independent, however, the $\tilde{e}_R$ pair production
cross-section is a function of the \mbox{GMSB}
 parameters due to the exchange of a
\nuno~in the \mbox{$t$--channel}. Therefore, in order to put limits on the
\selr~mass, the aforementioned region of the \mbox{GMSB} parameter space 
was scanned and, for each selectron mass, the smallest theoretical production
cross-section was chosen for comparison with the experimental limits. 
For gra\-vi\-ti\-no masses below a few\eVcc, the experimental limits are the
ones corresponding to the search for selectrons in MSUGRA models.
%Therefore, within the co-NLSP scenario, the impact parameter
%search and the se\-con\-da\-ry vertex search allow for the exclusion of
%$\tilde{\mu}_R$ masses  
%below 91\GeVcc\ for gravitino masses between 40 and 200\eVcc,
%and $\tilde{e}_R$ masses below 70\GeVcc~for $m_{\tilde{G}}$ between 15 and 20\eVcc.
%For \mgrav\ below a few\eVcc, $m_{\tilde{\mu}_R} >$ 80\GeVcc\
%was excluded by the search for \smur\ in gravity mediated
%models~\cite{slep_200}. On the other hand, the selectron mass lower 
%limit was in this case 57\GeVcc.

Assuming mass degeneracy between the sleptons, 
(Figure~\ref{fig:excl_202}-d), these searches 
extended the limit $\tilde{l}_R >$ 80\GeVcc~set by MSUGRA 
searches~\cite{slep189} for 
very short NLSP lifetimes, up to $m_{\tilde{G}}$= 700\eVcc. 
For the MSUGRA case no lepton combination exists, so the best limit from
the $\tilde{\mu}_R$ has been used.
The maximum excluded value
of $m_{\tilde{l}_R}$= 94\GeVcc~was achieved for $m_{\tilde{G}}$= 200\eVcc. For
$m_{\tilde{G}} >$ 700\eVcc~the best lower mass limit was set by the stable heavy
lepton search. 
%exclude at 95\% CL 
%$\tilde{l}_R$\ masses below 93.5\GeVcc\ for $\tilde{G}$ masses 
%between 30 and 250\eVcc.
%larger than 40\eVcc.
%For very short lifetimes only
%$\tilde{\mu}_R$ was considered since it gives the best limit that can be
%achieved in absence of slepton combination. 
%For $\tilde{G}$ larger than 1000\eVcc\
%the limit was 93\GeVcc, obtained from the
%stable heavy lepton search~\cite{heavyparticles-conf}. 
$\tilde{l}_R$\ masses below 35\GeVcc\ were excluded by LEP~1 data~\cite{lep1ex}.
In the case of $\tilde{l}_R$\ degeneracy, this limit improved to 41\GeVcc.

\subsection{Chargino pair production}

%Figure~\ref{fig:xsec} 
%shows, as an example, the 95\% CL upper limit on the chargino pair 
%production cross-section at $\sqrt{s} = 202$~GeV as a
%function of $m_{\tilde{\chi}_1^+}$\ and $m_{\tilde{l}_R}$\ 
%after combining the results of the searches at lower energies
%with the maximum likelihood ratio method~\cite{Read} for 
%$m_{\tilde{G}}$ = 100 \eVcc. 
The limits on the chargino pair production cross-section 
were used to exclude areas within the 
($m_{\tilde{\chi}_1^+},m_{\tilde{l}}$) plane for different 
domains of the gravitino mass combining results from 
all the centre-of-mass energies from 183~GeV to 202~GeV.
Figure~\ref{fig:mass} shows the regions excluded at 
95\% CL in the 
($m_{\tilde{\chi}_1^+}$,$m_{\tilde{\tau}_1}$) plane (a) and 
($m_{\tilde{\chi}_1^+}$,$m_{\tilde{l}_R}$) plane (b).
The positive-slope area is excluded for all gravitino masses. The 
negative-slope area is only excluded 
\mbox{for $m_{\tilde{G}} \gtrsim$ 100\eVcc}~
and the brick area for $m_{\tilde{G}} \gtrsim$ 1\keVcc.
The areas below $m_{\tilde{\tau}_1}$=~73\GeVcc\ (Figure~\ref{fig:mass}
(a)) and below $m_{\tilde{l}_R}$=~80\GeVcc\
(Figure~\ref{fig:mass} (b)), are excluded
by the direct search for slepton pair production in MSUGRA
models~\cite{slep189}. 
The area of $\Delta m  \leq  0.3$\GeVcc\ is not excluded because 
in this region the charginos do not decay mainly 
to \stuno\ and $\nu_{\tau}$, but to W and $\tilde{G}$.
Thus, if 
$\Delta m\! \geq\! 0.3$\GeVcc, the chargino mass   
limits are 95.2, 96.8 and 99\GeVcc\ for $m_{\tilde{G}}=$ 1, 100 and 
1000\eVcc\ respectively, in the ${\tilde{\tau}_1}$ NLSP scenario.
In the sleptons co-NLSP scenario the limits are 95.2, 98.6 and 98.6\GeVcc\
for \mbox{$m_{\tilde{G}}=$ 1}, 100 and 1000\eVcc\ respectively.
The limit at $m_{\tilde{G}}=$ 1\eVcc\ is also valid for 
smaller masses of the gravitino, because they lead to the same final state 
topologies. The same argument is true for 
$m_{\tilde{G}} \gtrsim$ 1\keVcc. The chargino mass limit 
decreases with decreasing $m_{\tilde{\tau}_1}$\   because 
in scenarios with gravitino LSP, small stau masses correspond to 
small sneutrino masses (both are proportional to $\Lambda$) 
and hence to smaller production 
%KH cross sections 
cross-sections due to the destructive interference between the 
%KH s- and t-channels.
$s$- and $t$-channels.
It should be noticed that within the parameter space covered by this work, 
the lightest chargino is at least 40\% heavier than the lightest neutralino. 
Thus, for gravitino masses up to \mbox{$\sim$ 1\eVcc}~the search for
neutralinos implies a model dependent lower limit
on the lightest chargino of 125\GeVcc. However, neutralinos were not directly
searched for in heavier gravitino mass regions, therefore, a model
dependent lower limit cannot be set in this case. Thus, the experimental lower
limit of 98.6\GeVcc\ remains valid.   

%These
%limits, which directly 
%reflect the efficiencies of the
%applied selections, can be understood as follows:
%
%\begin{description}
%\item[$m_{\tilde{G}}=1$\eVcc :] 
%The efficiency of this analysis depends mainly on the mass of the 
%chargino. To smaller chargino masses correspond bigger event missing energies,
%and bigger efficiencies.
%flat over the kinematically 
%allowed  region. Thus, the main effect on the production cross section limit 
%is given by the 211.9~pb$^{-1}$\ used for the region 
%$m_{\tilde{\chi}^+_1} < 91.5$\GeVcc, and the 158~pb$^{-1}$\ for 
%$m_{\tilde{\chi}^+_1} > 91.5$\GeVcc.
%
%\item[$m_{\tilde{G}}=100$\eVcc :] The map of efficiencies is the result of 
%the convolution of two factors. First, 
%larger stau masses imply a 
%smaller lifetime, and hence a smaller efficiency. Second, 
%a larger chargino mass leads
%to smaller stau momenta, and to smaller 
%decay lengths.
%
%\item[$m_{\tilde{G}}=1000$\eVcc :] In this case, the map of efficiencies 
%is mainly 
%affected by the momentum of the stau, because the method used to identify 
%heavy stable particles relies on the lack of Cherenkov radiation 
%in DELPHI's RICH detectors. 
%To remove SM backgrounds, low momentum particles 
%are removed, thus reducing the efficiency for higher chargino masses, 
%especially in the region of small  $\Delta m$.
%\end{description}

\subsection{Heavy stable charged particle pair production} 
The results presented in section~\ref{heavystable} were combined with 
previous DELPHI results in this channel,
and cross-section limits were derived
as indicated in Figure~\ref{limit}. From the intersection points with
the predicted cross-sections for smuon or staus in the MSSM, left(right)
handed smuons and staus can be excluded up to masses
of 94.0(93.5)~GeV/$c^2$ at 95\%CL. No limits are given on selectrons here,
because the cross-section can be highly suppressed by an additional
t-channel sneutrino exchange contribution.

\subsection{Limits on the GMSB parameter space} 

Finally, all these results can be combined to produce exclusion plots 
within the ($\tan\beta, \Lambda$) space. 
The corresponding pa\-ra\-me\-ter space was scanned as follows:
$1\leq n \leq 4$, $5\ {\rm TeV}\leq\Lambda\leq 90\ {\rm TeV}$, 
$1.1\leq M/\Lambda \leq 10^9$, $1.1\leq \tan\beta\leq 50$, and 
$sign(\mu) = \pm 1 $.
As an example,  Figure~\ref{fig:lambda} shows the zones
 excluded for $n=$1 to 4 for $m_{\tilde{G}} \leq 1$\eVcc , which corresponds to
the NLSP decaying at the main vertex. 
The shaded areas are excluded. The areas below 
the dashed lines contain points of the GMSB parameter space with \nuno~NLSP.
The areas to the right (above for $n = 1$) of the dashed-dotted lines 
contain points of the GMSB parameter space where sleptons are the NLSP. 
It can be seen that the region of slepton NLSP increases with $n$. 
The contrary occurs to the region of neutralino NLSP. 
A limit could be set for the variable $\Lambda$\ at 17.5~TeV.

%%%%%%%%%%%%%%%%%%%%%%%%%%%%%%
\section{Summary}
Lightest neutralino, slepton and chargino pair production 
were searched for in the 
context of light gravitino models. Two possibilities
were explored: the $\tilde{\tau}_1$ NLSP  and the sleptons co-NLSP
scenarios.  No evidence for signal production was found.
Hence, the DELPHI collaboration 
sets lower limits at 95\% CL for the mass of 
the $\tilde{\chi}_1^0$ at 86\GeVcc\ if \mbox{$m_{\tilde{G}}~\lesssim$ 1\eVcc}, 
and lower mass limits for the sleptons in all the gravitino mass range.
%for the mass of 
%the $\tilde{\tau}_R$ at 88\GeVcc, for the mass of the $\tilde{\mu}_R$ at
%91\GeVcc, and finally for the mass of the  
%$\tilde{l}_R$ at 93.5\GeVcc~for $m_{\tilde{G}}$ in the range between 
%20 and 300\eVcc. For $\tilde{e}_R$ the mass limit is 70\GeVcc~
%for $m_{\tilde{G}}$ in the range between 20 and 110\eVcc. 
The limit on the chargino mass is 95.2\GeVcc~for all $m_{\tilde{G}}$ in both
scenarios, $\tilde{\tau}_1$ NLSP  and the sleptons co-NLSP.

Finally, mass limits for heavy stable charged particles were also derived
within the MSSM. For these particles the DELPHI collaboration sets lower mass
limits at 95\% CL for the left (right) handed sleptons at 94.0 (93.5)\GeVcc. 
%         Modified on 04-06-1999 by dimartino
%-------------------------------------------------------------------
\subsection*{Acknowledgements}
\vskip 3 mm
 We are greatly indebted to our technical 
collaborators, to the members of the CERN-SL Division for the excellent 
performance of the LEP collider, and to the funding agencies for their
support in building and operating the DELPHI detector.\\
We acknowledge in particular the support of \\
Austrian Federal Ministry of Education, Science and Culture,
GZ 616.364/2-III/2a/98, \\
FNRS--FWO, Flanders Institute to encourage scientific and technological 
research in the industry (IWT), Belgium,  \\
FINEP, CNPq, CAPES, FUJB and FAPERJ, Brazil, \\
Czech Ministry of Industry and Trade, GA CR 202/96/0450 and GA AVCR A1010521,\\
Commission of the European Communities (DG XII), \\
Direction des Sciences de la Mati$\grave{\mbox{\rm e}}$re, CEA, France, \\
Bundesministerium f$\ddot{\mbox{\rm u}}$r Bildung, Wissenschaft, Forschung 
und Technologie, Germany,\\
General Secretariat for Research and Technology, Greece, \\
National Science Foundation (NWO) and Foundation for Research on Matter (FOM),
The Netherlands, \\
Norwegian Research Council,  \\
State Committee for Scientific Research, Poland, 2P03B06015, 2P03B11116 and
SPUB/P03/DZ3/99, \\
JNICT--Junta Nacional de Investiga\c{c}\~{a}o Cient\'{\i}fica 
e Tecnol$\acute{\mbox{\rm o}}$gica, Portugal, \\
Vedecka grantova agentura MS SR, Slovakia, Nr. 95/5195/134, \\
Ministry of Science and Technology of the Republic of Slovenia, \\
CICYT, Spain, AEN96--1661 and AEN96-1681,  \\
The Swedish Natural Science Research Council,      \\
Particle Physics and Astronomy Research Council, UK, \\
Department of Energy, USA, DE--FG02--94ER40817. \\
%=========================================================================%
%=========================================================================%
\newpage
%
%%%%%%%%%%%%%%%%%%%%%%%%%%%%%%%%%%%%%%%%%%
%% BIBLIO
%%%%%%%%%%%%%%%%%%%%%%%%%%%%%%%%%%%%%%%%%%

\newpage

\clearpage

%%%%%%%%%%%%%%%%%%%%%%%%%%%%%%
%%   Figures
%%%%%%%%%%%%%%%%%%%%%%%%%%%%%%%%%%%%%%%%%%
%\begin{figure}[hbpt]
%\vspace{-1cm}
%\centerline{\epsfxsize=16.0cm \epsfysize=21.0cm \epsfbox{angles_e200.eps}} 
%\vspace{-0.5cm}
%   \caption[]
%{(a) Angle between the directions defined by the hadronic vertex and the
%   reconstructed vertex, (b) angle between the tracks of the kink, and (c)
%angle between the electromagnetic shower
%  and the direction defined by the difference between the momenta of the
%   $\tilde{\tau}_1$ and its associated $\tau$, defined at the crossing
%  point
% for real data (dots), expected Standard Model background (cross-hatched 
%histogram) and simulated signal for $m_{\tilde{\tau}_1} = 60$\GeVcc\
%decaying with a mean 
% distance of 50~cm (blank histogram). Events that do not have hadronic 
%interactions are not included in (a), and events without electromagnetic 
%showers are not included in (b). All the samples are at 200 GeV
%   centre-of-mass energy.
%The arrows indicate the selection criteria imposed.}
%  \label{fig:grav:kinks_BG}
%\end{figure}
%%%%%%%%%%%%%%%%%%%%%%%%%%%%%%%%%%%%%%%%%%%
%\begin{figure}[hbpt]
%\vspace{-1cm}
%\centerline{\epsfxsize=16.0cm \epsfbox{ip.eps}} 
%   \caption[]
%{$\sqrt{b_1^2 + b_2^2}$ distribution for data (dots) and simulated  
%backgrounds (histogram) after all other cuts applied by the small impact
%parameter search.}
%  \label{fig:ip-data-mc}
%\end{figure}
%
\begin{figure}[htb]
\begin{center}
\mbox{\epsfxsize17.0cm 
\epsffile{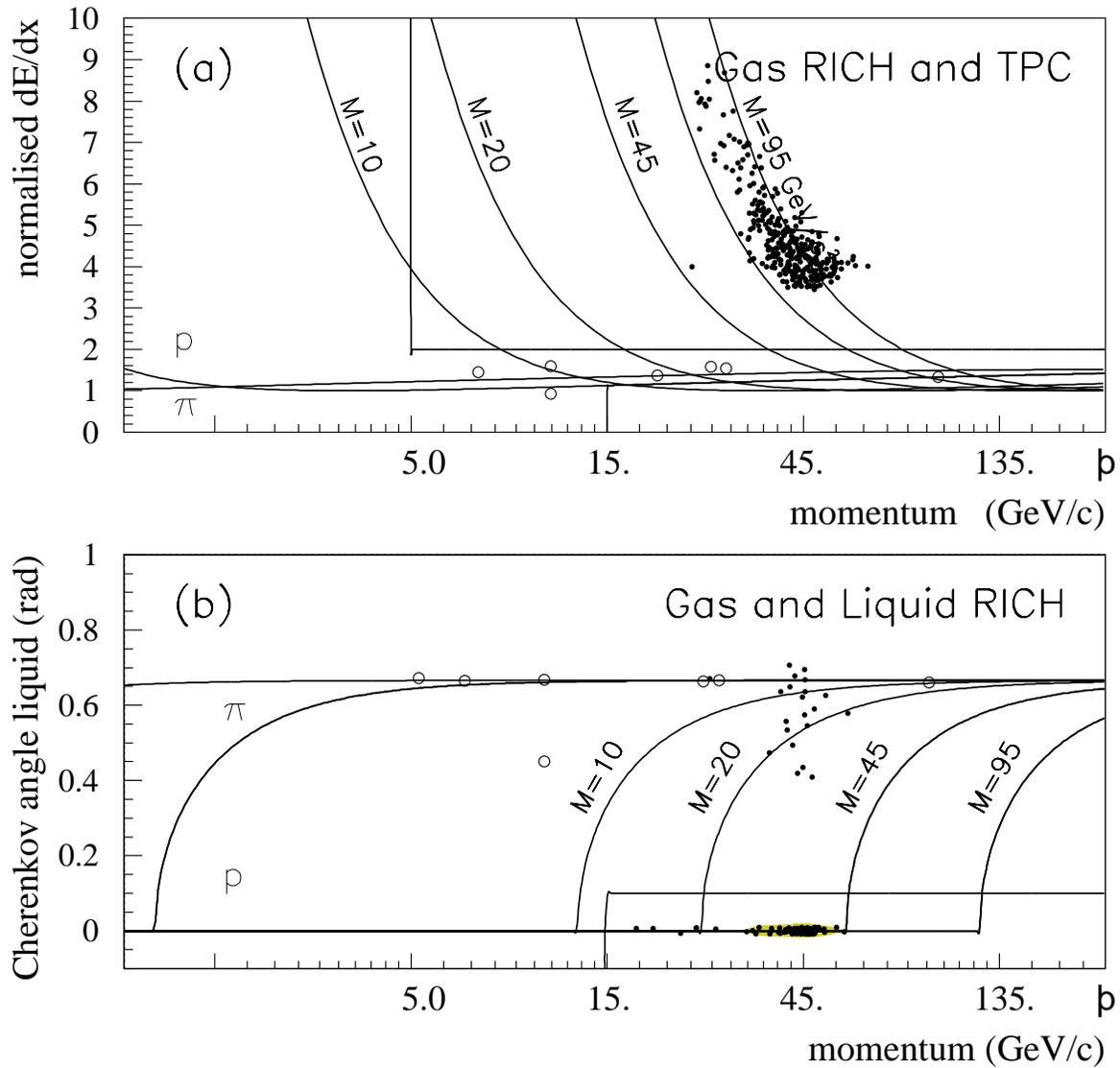} }
\end{center}
\vspace*{0.cm}
\caption{
(a) Normalised energy loss as a function of the momentum 
 after the gas veto for the 202 GeV data.
(b) Measured Cherenkov angle in the liquid radiator 
as a function of the momentum after the gas veto:
if four photons or less were observed 
in the liquid radiator, the Cherenkov angle was set equal to zero.
The rectangular areas in (a) indicate selections (1) and (2),
and that in (b) shows selection (3). The selection criteria are
explained in the text. Open circles are data. The small filled circles
indicate the expectation for a 90\GeVcc~mass signal with charge $\pm$e,
resulting in a large dE/dx (upper
plot) and no photons (except for a few accidental rings) in the liquid
Cherenkov counter (lower plot). The solid lines with a mass signal value
 indicate the expectation for heavy stable sleptons.}
\label{rare202}
\end{figure}

%%%%%%%%%%%%%%%%%%%%%%%%%%%%%%%%%%%%%%%%%%%%%%%%%%%%%%%%%%%%%%%%%%%%%%%%%%%%%
\begin{figure}[htbp]\centering
%\vspace{-1.cm}
\centerline{\epsfxsize=14.0cm \epsffile{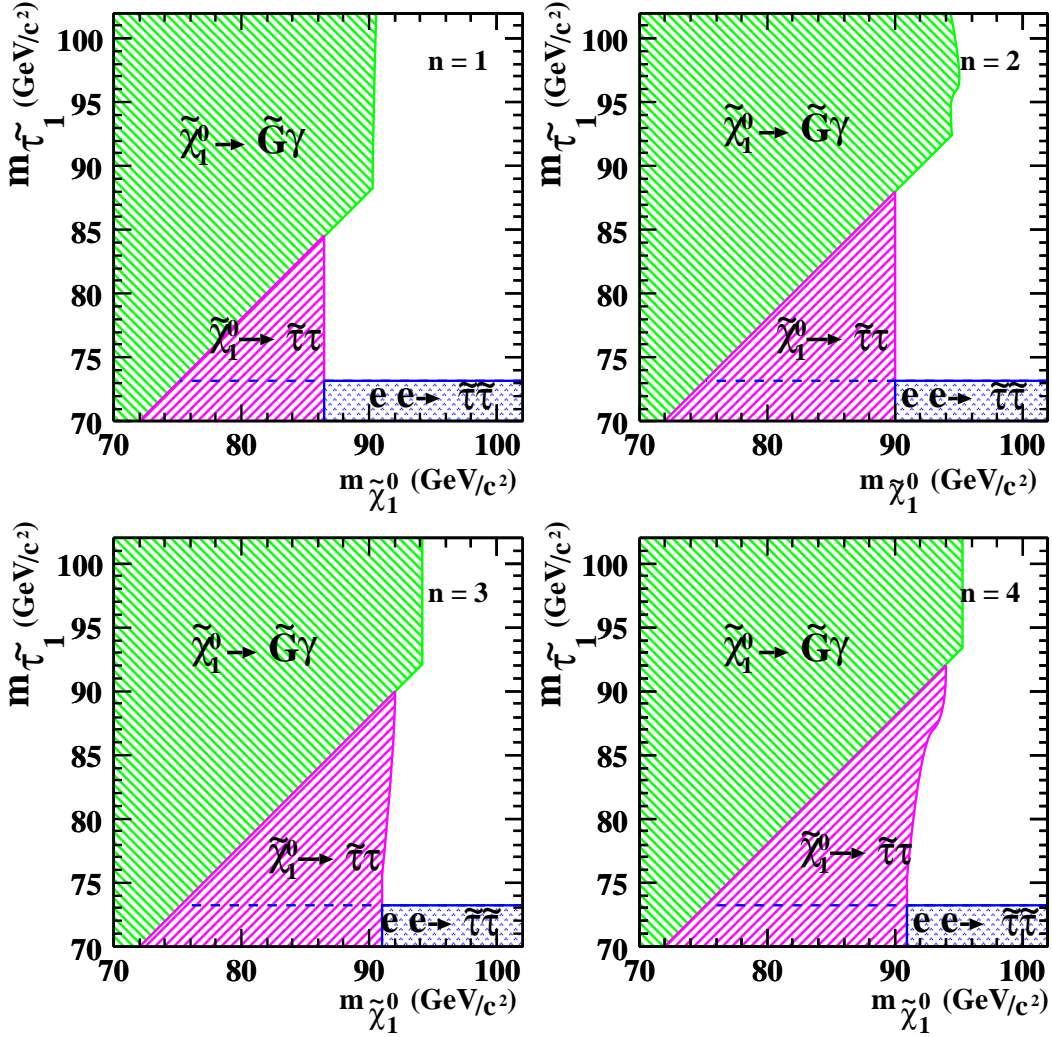}}
\caption[]{Areas excluded at 95\% CL with $m_{\tilde{G}}< 1$ \eVcc~ in the 
$m_{\tilde{\chi}_1^0}$\ vs $m_{\tilde{\tau}_1}$ plane for $n = $1 to 4, using
all data from 161 GeV to 202 GeV centre-of-mass energies. 
The positive-slope dashed area is excluded by this analysis.
The negative-slope dashed
area is excluded by the search for
$\tilde{\chi}^0_1\rightarrow \gamma \tilde{G}$,
 and the point-hatched area by the direct search for stau pair 
production in the 
MSUGRA framework.}
\label{fig:masses}
\end{figure}

%%%%%%%%%%%%%%%%%%%%%%%%%%%%%%%%%%%%%%%%%%%%%%%%%%%%%%%%%%%%%%%%%%%%%%%%%%%%%
%\begin{figure}[htbp]
%\centerline{\epsfxsize=16.0cm \epsfbox{excl_stau_r_202.eps}}
%  \caption[]{ 
%    Exclusion region in the 
%    ($m_{\tilde{G}}$,$m_{\tilde{\tau}_R}$) plane (a) and 
%    ($m_{\tilde{G}}$,$m_{\tilde{\tau}_1}$) plane (b) 
%    at 95\%~C.L. for the present analysis combined 
%    with the stable heavy
%    lepton search and the search for $\tilde{\tau}_1$ within MSUGRA models, 
%    using all LEP-2
%    data up to 202 GeV. 
%    The positive-slope hatched area shows the region excluded 
%    by the impact parameter and secondary vertex searches.
%} 
%  \label{fig:stau1-mass}
%\end{figure}

%%%%%%%%%%%%%%%%%%%%%%%%%%%%%%%%%%%%%%%%%%%%%%%%%%%%%%%%%%%%%%%%%%%%%%%%%%%%%
\begin{figure}[H]
\begin{tabular}{cc}
\epsfxsize=8.cm\epsfysize=8.cm\epsfbox{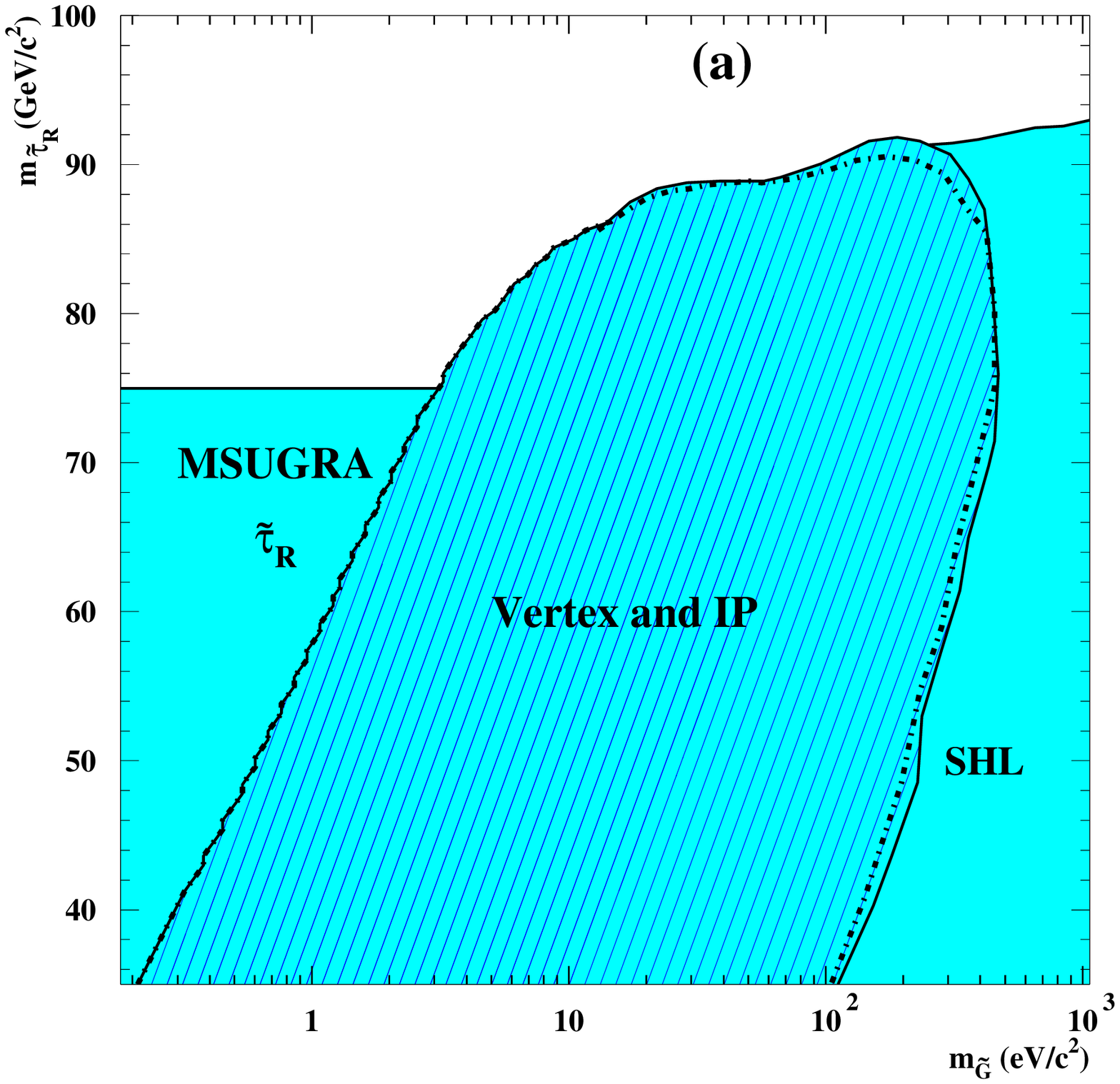} & 
\epsfxsize=8.cm\epsfysize=8.cm\epsfbox{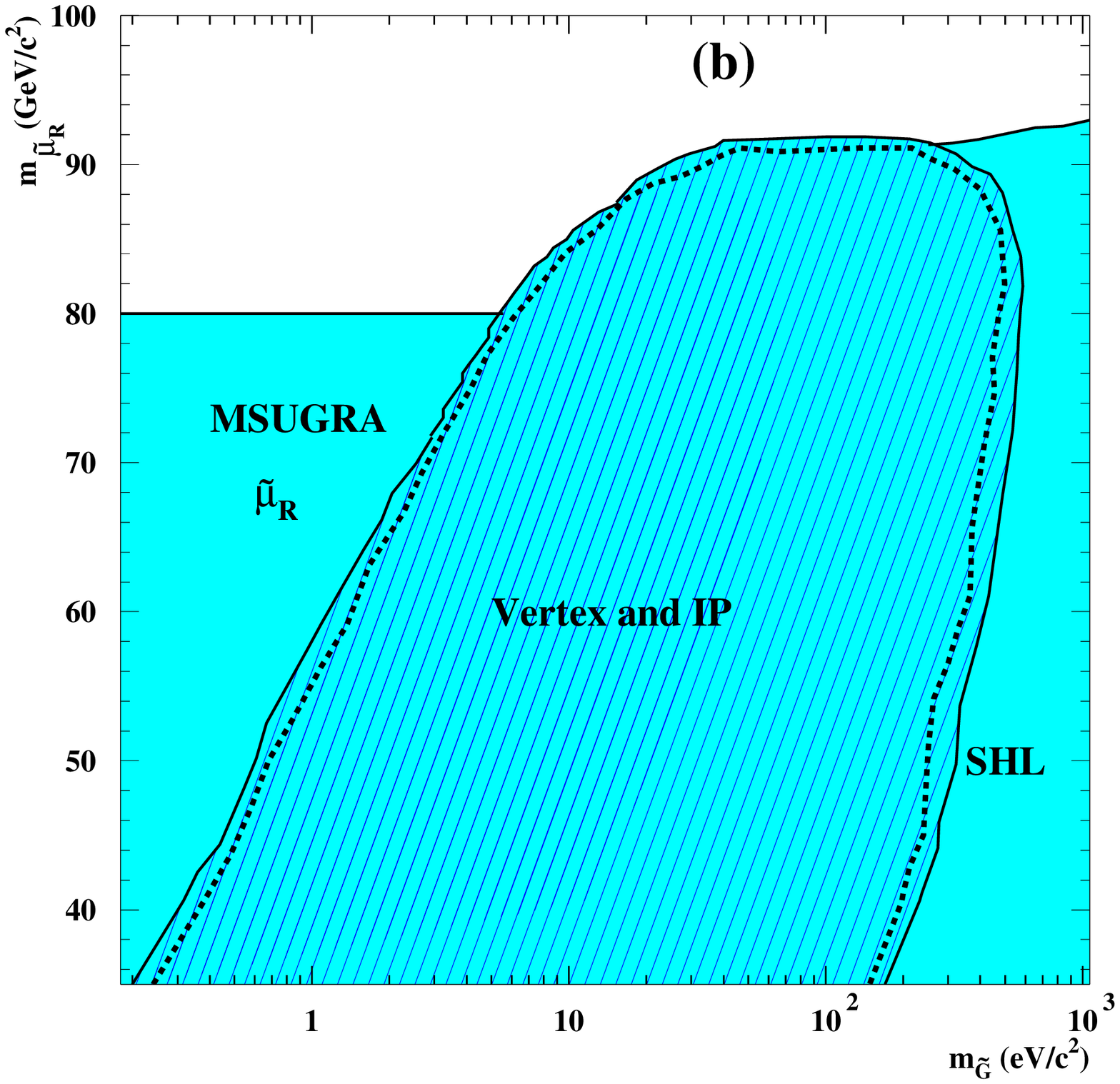} \\ 
\epsfxsize=8.cm\epsfysize=8.cm\epsfbox{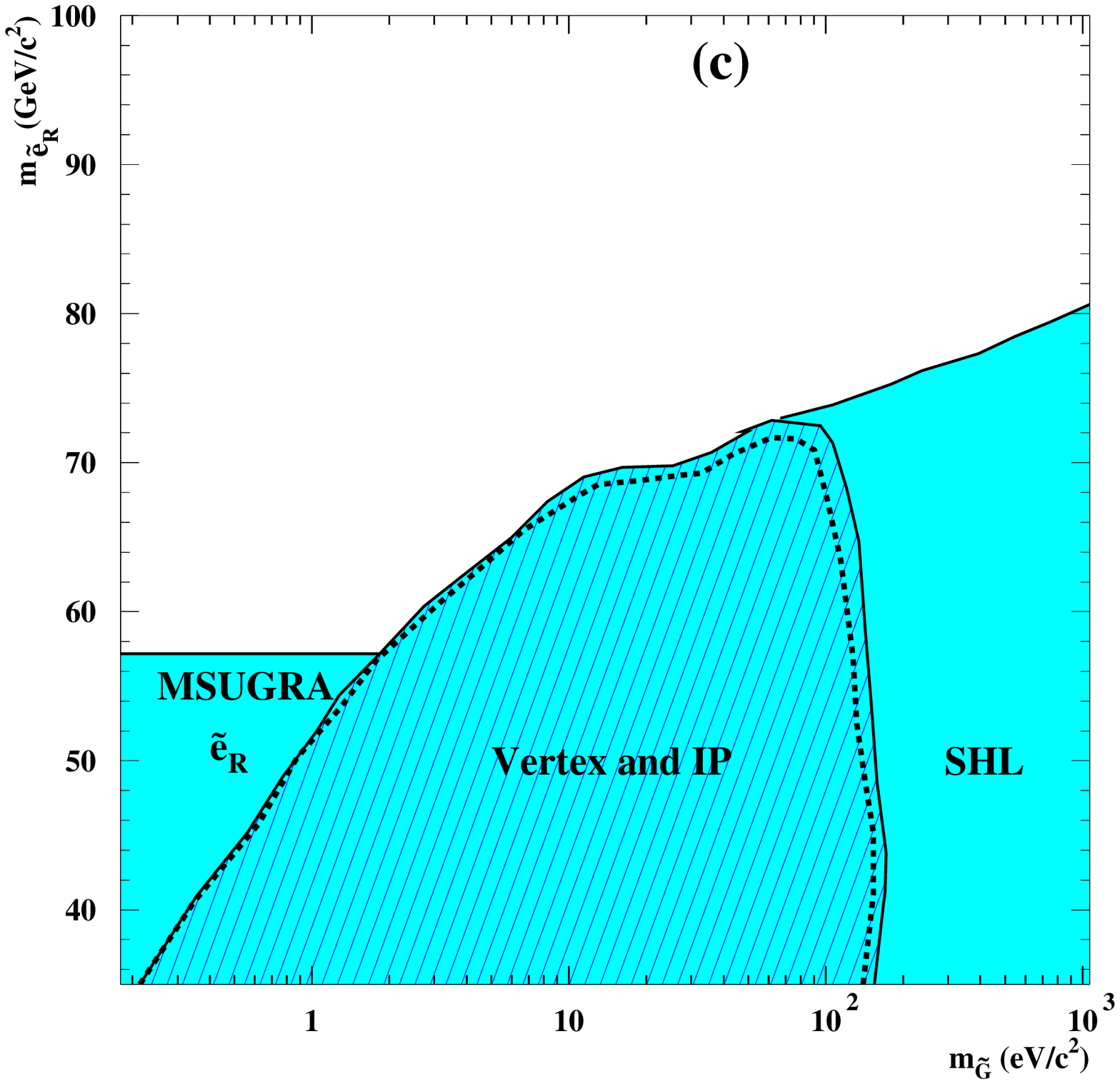} &
\epsfxsize=8.cm\epsfysize=8.cm\epsfbox{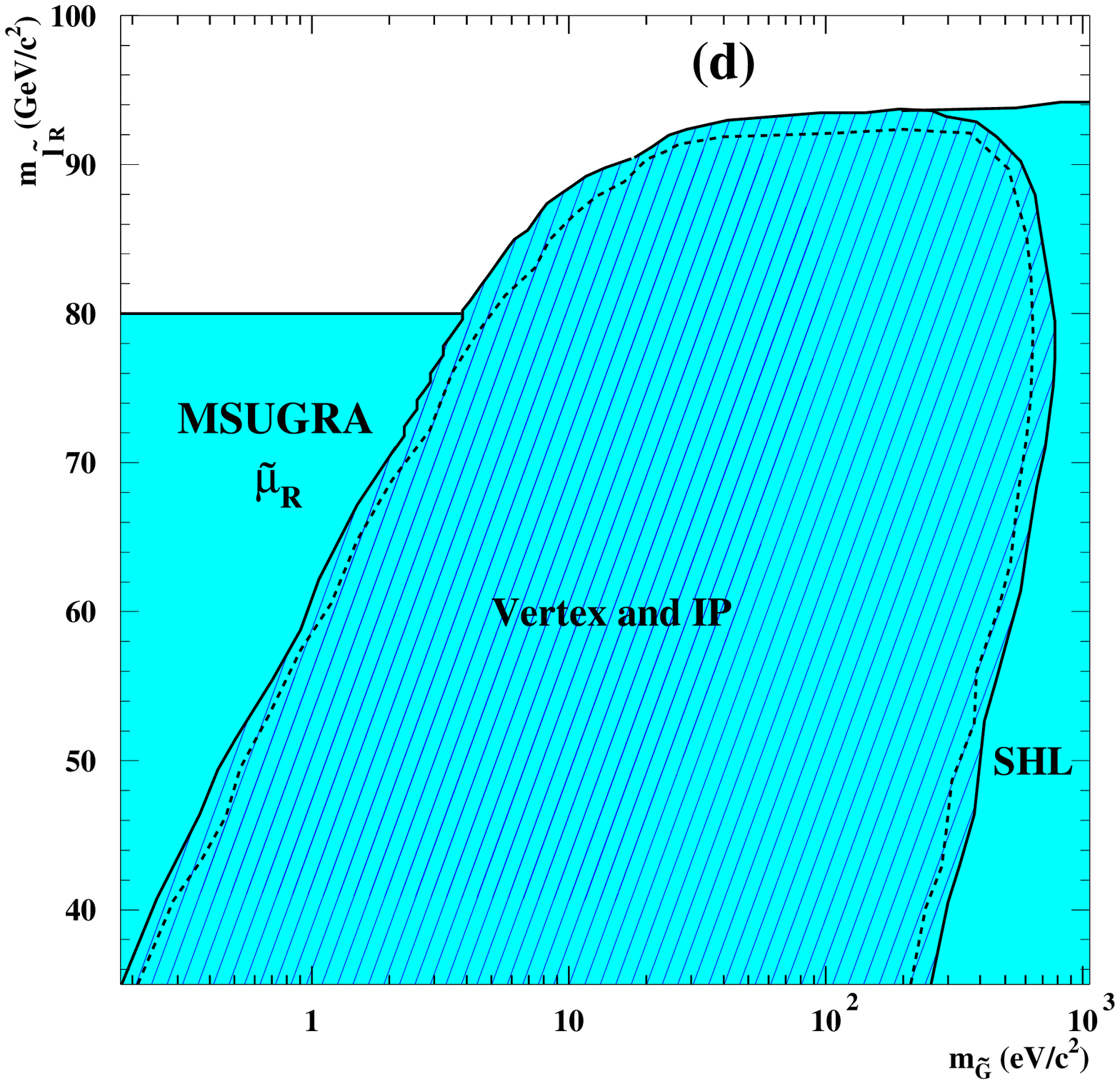} \\ 
\end{tabular}
  \caption[]{ 
    Exclusion regions in the 
    ($m_{\tilde{G}}$,$m_{\tilde{\tau}_R}$) (a),
    ($m_{\tilde{G}}$,$m_{\tilde{\mu}_R}$) (b),
    ($m_{\tilde{G}}$,$m_{\tilde{e}_R}$) (c)
    ($m_{\tilde{G}}$,$m_{\tilde{l}_R}$) (d)
     planes at 95\%~CL for the present analyses combined 
    with the Stable Heavy
    Lepton (SHL) search and the search for $\tilde{l}$ in gravity mediated
    models (MSUGRA), using all DELPHI data from 130 GeV to 202 GeV
    centre-of-mass energies. 
    The positive-slope hatched area shows the region excluded 
    by the combination
    of the impact parameter and secondary vertex searches.
    The dashed line shows the expected limits.
} 
  \label{fig:excl_202}
\end{figure}

%\begin{figure}[htbp]\centering
%\epsfxsize=16.0cm
%\centerline{\epsffile{xsec_cha_202.eps}}
%\caption{Limits in picobarns on the lightest chargino pair production
%  cross-section at 95\% CL for 202 GeV centre-of-mass energy. Limits are
%    shown as functions of $m_{\tilde{\chi}_1^+}$ and $m_{\tilde{l}_R}$ for
%    $m_{\tilde{G}}=1$00\eVcc.} 
%\label{fig:xsec}
%\end{figure}

\begin{figure}[htbp]
\epsfxsize=16.0cm
\centerline{\epsffile{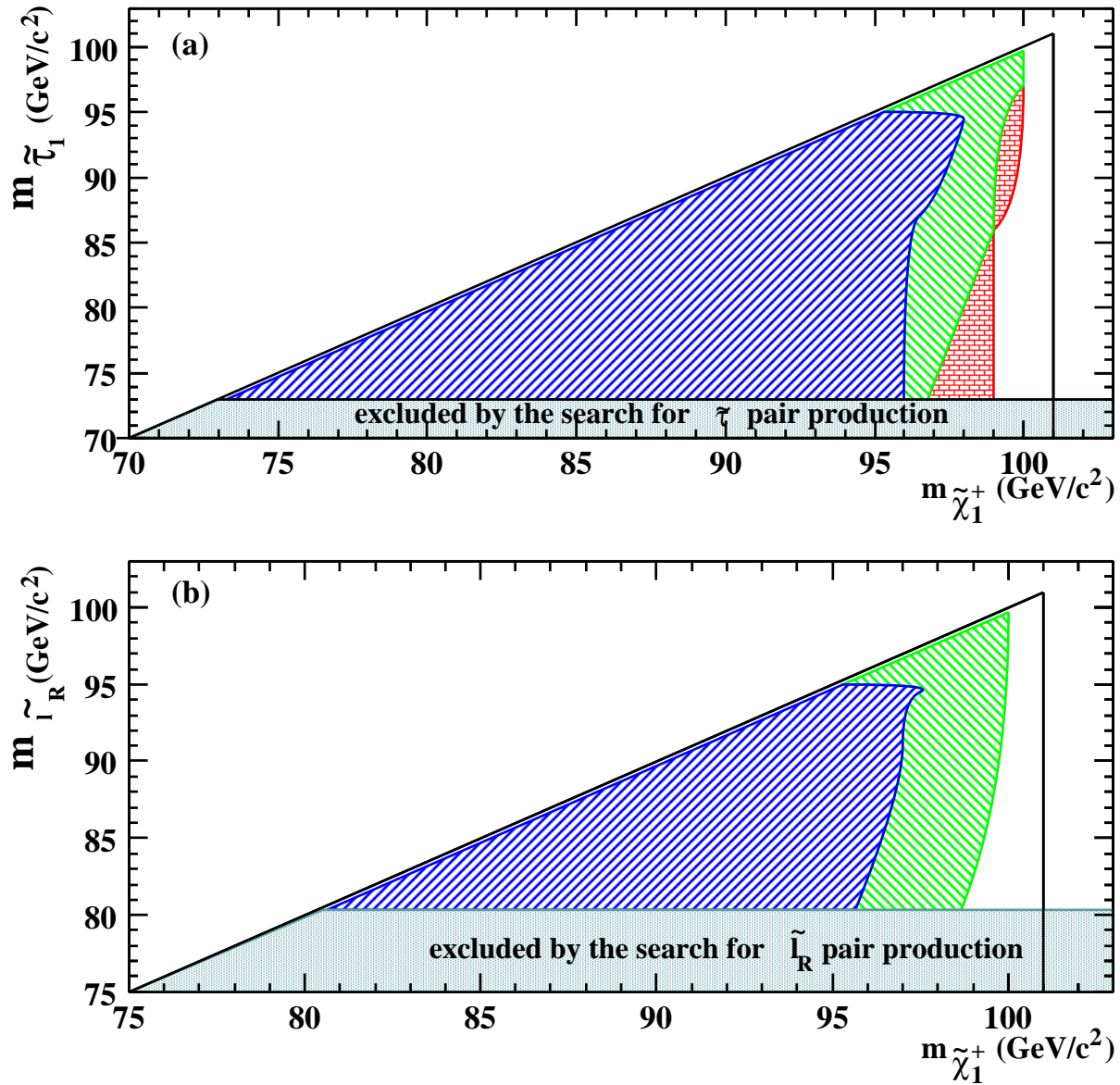}}
\caption[]{Areas excluded at 95\% CL in the 
($m_{\tilde{\chi}_1^+}$,$m_{\tilde{\tau}_1}$) plane (a) and
($m_{\tilde{\chi}_1^+}$,$m_{\tilde{l}_R}$) plane (b) using all DELPHI
data from 183 GeV to 202 GeV centre-of-mass energies.
The positive-slope area is excluded for all $m_{\tilde{G}}$.
The negative-slope area is excluded only for  
$m_{\tilde{G}} \geq 1$00\eVcc~and the brick area for $m_{\tilde{G}} \gtrsim$
1\keVcc. The grey area is excluded 
by the direct search for slepton pair production within MSUGRA models.}
\label{fig:mass}
\end{figure}

\begin{figure}[htb]
\begin{center}
\epsfxsize=8.0cm\centerline{\epsffile{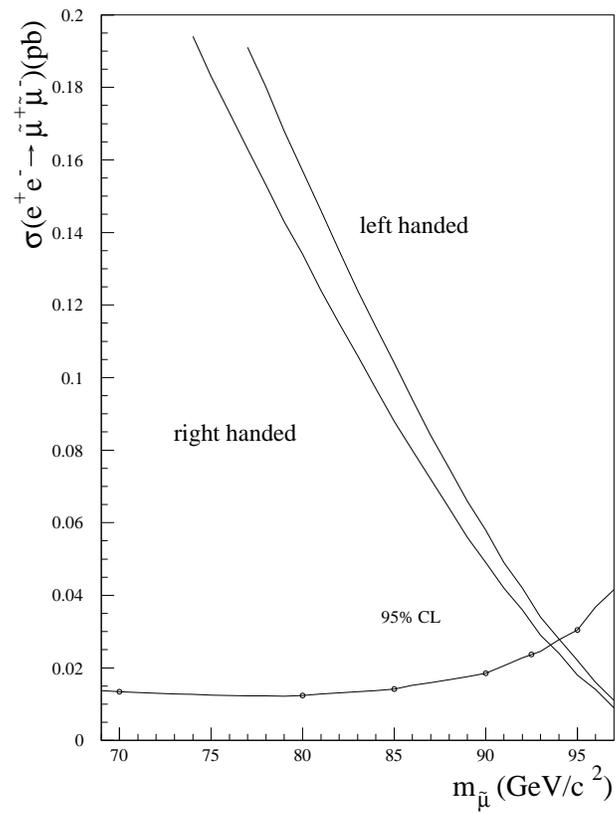}}
\end{center}
\vspace*{0.cm}
\caption{Predicted production cross-section for left and right handed
stable smuons (staus) as a function of the particle mass. The cross-section
limit indicated in the figure has been derived using all DELPHI data
between 130 and 202~GeV.}
\label{limit}
\end{figure}

\begin{figure}[htbp]
\epsfxsize=16.0cm
\centerline{\epsffile{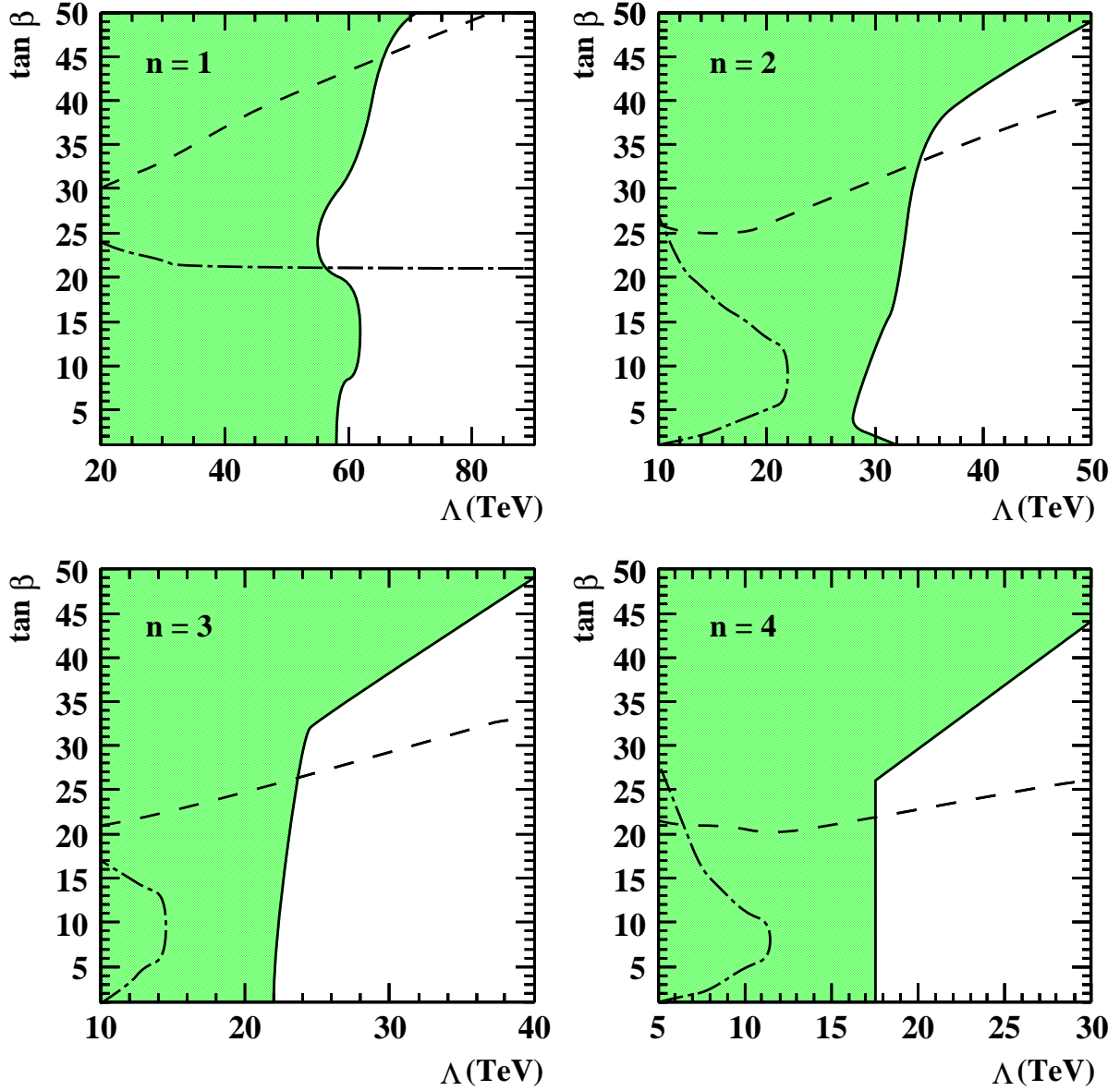}}
\caption{Shaded areas in the 
($\tan\beta , \Lambda$) plane are excluded at 95\% CL.
The areas below 
the dashed lines contain points of the GMSB parameter space with \nuno~NLSP.
The areas to the right (above for $n = 1$) of the dashed-dotted lines 
contain points of the GMSB parameter space were sleptons are the NLSP.}
\label{fig:lambda}
\end{figure}

\end{document}